%% file: imp_spec_mwt.tex
\documentclass[a4paper,12pt]{article}

\usepackage{amsmath,amssymb,amsfonts}
\usepackage[pdftex]{graphicx,color}
\usepackage{subfig}
\usepackage[toc,page]{appendix}
\usepackage{flafter}
\usepackage[section]{placeins}
\usepackage{slashed}
\usepackage{fullpage}
\usepackage{hyperref}
\numberwithin{equation}{section}
\usepackage{enumerate}
\usepackage{multirow}

\newcommand\email[1]{{\tt\href{mailto:#1}{#1}}}

\begin{document}

\begin{titlepage}

\setcounter{page}{0}

\vspace*{-2cm}

\begin{flushright}
{Edinburgh 2011/15}, {WUB/11-02}, {CP3-Origins-2011-11}, {CERN-PH-TH/2011-090}
\end{flushright}

\vspace{0.6cm}

\begin{center}
{\Large \bf Improved Lattice Spectroscopy of Minimal Walking Technicolor} \\ 

\vskip 0.8cm

{\bf Francis Bursa}\\
{\sl
Jesus College,
University of Cambridge,
Cambridge,
England\\
E-mail: \email{f.bursa@damtp.cam.ac.uk}
}

\vskip .2cm

{\bf Luigi Del Debbio}\\
{\sl
SUPA, School of Physics and Astronomy,
University of Edinburgh,
Edinburgh,
Scotland\\
E-mail: \email{luigi.del.debbio@ed.ac.uk}
}

\vskip .2cm

{\bf David Henty}\\
{\sl
EPCC, University of Edinburgh, Edinburgh, Scotland\\
E-mail: \email{d.henty@epcc.ed.ac.uk}
}

\vskip .2cm

{\bf Eoin Kerrane}\\
{\sl
SUPA, School of Physics and Astronomy,
University of Edinburgh,
Edinburgh,
Scotland\\
E-mail: \email{eoin.kerrane@ed.ac.uk}
}

\vskip .2cm

{\bf Biagio Lucini}\\
{\sl 
College of Science,
Swansea University,
Swansea,
Wales\\
E-mail: \email{b.lucini@swansea.ac.uk}
}

\vskip .2cm

{\bf Agostino Patella}\\
{\sl 
CERN,
Geneva,
Switzerland\\
E-mail: \email{agostino.patella@cern.ch}
}

\vskip .2cm

{\bf Claudio Pica}\\
{\sl 
CP$^\mathit{3}$-Origins \& IMADA,
University of Southern Denmark,
Odense,
Denmark\\
E-mail: \email{pica@cp3.sdu.dk}
}

\vskip .2cm

{\bf Thomas Pickup}\\
{\sl Rudolf Peierls Centre for Theoretical Physics,
 University of Oxford, 
Oxford,
 England\\
E-mail: \email{pickup@thphys.ox.ac.uk}
}

\vskip .2cm

{\bf Antonio Rago}\\
{\sl 
Department of Physics,
Bergische Universit\"at Wuppertal,
Wuppertal,
Germany\\
E-mail: \email{rago@physik.uni-wuppertal.de}
}

\vskip .6cm

\end{center}

\begin{abstract}
We present a numerical study of spectroscopic observables in the $SU(2)$ gauge theory with two adjoint fermions using improved source and sink operators. We compare in detail our improved results with previous  determinations of masses that used point sources and sinks and we investigate possible systematic effects in both cases. Such comparison enables us to clearly assess the impact of a short temporal extent on the physical picture, and to investigate some effects due to the finite spatial box. While confirming the IR-conformal behaviour of the theory, our investigation shows that in order to make firm quantitative predictions, a better handle on finite size effects is needed.
\end{abstract}

\vfill

\end{titlepage}

\section{Introduction}

A new strongly interacting theory~\cite{Weinberg:1975gm,Susskind:1978ms}
with an approximate~\cite{Holdom:1984sk,Yamawaki:1985zg,Appelquist:1986an} or exact~\cite{Luty:2004ye}
infra-red (IR) fixed point is an appealing possibility for explaining electroweak symmetry
breaking. This framework, known as Technicolor, has been reviewed recently in
e.g.~\cite{Hill:2002ap,Sannino:2009za,Piai:2010ma}. Technicolor theories
are inherently non-perturbative and therefore require adequate tools to study their strong dynamics. Theories with conformal or near-conformal dynamics can be exposed in
the context of the gauge-string duality~\cite{Nunez:2008wi,Elander:2009pk}.
However, in addition to the wanted fermion and gauge boson degrees of
freedom, field theory duals of string theories in general contain
extra scalar fields. A possible {\em ab initio} approach relies on numerical simulations
of candidate Technicolor theories discretised on a spacetime lattice
(see e.g.~\cite{DeGrand:2010ba,DelDebbio:2010zz} for recent reviews). 

One could generate an infra-red fixed point in a gauge
theory by adding a low number of fermion flavours in higher
gauge representations to a gauge theory with a low number of
colours. The minimal vector-like gauge theory with this property, termed minimal walking technicolor (MWT),
has gauge group $SU(2)$ and two flavours of Dirac fermions in the adjoint
representation~\cite{Dietrich:2006cm}.  Some recent lattice studies of MWT~\cite{Hietanen:2009az,Bursa:2009we,DeGrand:2011qd} have
attempted to identify a near-conformal behaviour directly from the
behaviour of the coupling and anomalous dimensions of the theory under
RG flow.
Others,
including this work, perform measurements of physical observables in
the theory and attempt to identify signals of
near-conformal dynamics from their
behaviour~\cite{Catterall:2007yx,DelDebbio:2008zf,Catterall:2008qk,Hietanen:2009zz,DelDebbio:2009fd,Catterall:2009sb,DelDebbio:2010hx,DelDebbio:2010hu}. All the
evidence accumulated so far for this
theory favours a conformal or near-conformal scenario and seems to
exclude standard confinement and chiral symmetry breaking behaviour. However, more
systematic studies need to be performed before the IR properties of the
theory can be determined with confidence.  

MWT with a non-zero fermion mass and defined in a finite volume, as simulated for practical
reasons on the lattice, cannot be conformal. If the chiral continuum theory possesses an
infra-red fixed point, the lattice results will be described by a
mass-deformed conformal gauge theory~\cite{DeGrand:2009mt,DelDebbio:2009fd,Lucini:2009an,DelDebbio:2010hu,DelDebbio:2010hx,DelDebbio:2010jy,DelDebbio:2010ze}.
In approaching a conformal limit, the theory respects the hyperscaling
property, whereby all spectral masses $M$ in the theory scale identically. They
must vanish in the limit of vanishing fermion mass $m$.
If the IR fixed point is approximate, the theory displays conformal behaviour for an intermediate range of
masses $m$ and crosses over to the confining and chiral symmetry
breaking behaviour in the chiral limit.

The standard way to extract masses from lattice simulations is to look
at the exponential decay of correlators of operators with the quantum
numbers of interest. For infinite separation between the source and sink
operator, the exponential decay is governed by the ground state mass in the
channel being explored. At finite time extent, this leading behaviour receives
corrections that are exponentially suppressed in the mass difference between
the ground state and the excitations. Underestimating the importance of these
corrections leads to systematic errors in the determination of the ground state
mass. In addition to the effects of the finite maximal separation between
the source and the sink (often referred to as finite temperature effects),
the finite spatial extension of the lattice can also give
sizeable corrections to the spectral masses.

The simplest source and sink observables to study for mesons are
fermion bilinears in which the two fermion fields are at the same
lattice point ({\em point sources}). These sources have been widely used
in previous investigations of the spectrum of MWT. However, the experience accumulated over 30
years of numerical studies in lattice QCD favours the use of {\em
extended sources},  which are gauge-invariant combinations of two fermion
fields at different points, engineered for reducing the contamination
from the excited states. In lattice QCD
masses extracted from correlators of extended sources prove to be
affected by smaller systematic errors. In this paper, we investigate
whether this proves to be the case also for MWT. Specifically,
we perform a study of mesonic observables extracted from extended
sources using the configurations presented in~\cite{DelDebbio:2008zf,DelDebbio:2009fd,DelDebbio:2010hu}. 
We explore a large set of schemes for building extended operators and we systematically analyse their efficiency for the computation of meson masses and decay
constants, comparing the results with results obtained using point sources. 
With this study, we expect to determine the size of
systematic uncertainties in current studies, which have as yet been largely
unexplored, and to assess their impact on the physical picture
emerging from the previous spectroscopical studies.  
Some of the results presented here have already appeared in Ref.~\cite{Kerrane:2010xq}.

The rest of the paper is organised as follows. In Sec.~\ref{sec:systematics} we describe the background to this study and briefly illustrate the effects of the
use
of different smearings on effective observables. Technical details on the smearing procedures and the resulting observables can be found in
Appendices~\ref{app:smear} and \ref{app:mesoncorr} respectively. In Sec.~\ref{sec:evaluate} we quantify the consequences of the smearing both for
autocorrelation times and quality of plateaux. A full set of results obtained using wall smearing are presented in Sec.~\ref{sec:results}, while in
Sec.~\ref{sec:finitevolume} we comment on the significant finite-volume effects highlighted by the smeared results. Appendices~\ref{app:results}~and~\ref{app:pull} list the numerical values of the quantities studied in this work. Finally, our conclusions are reported in Sect.~\ref{sec:conclusions}.

\section{Systematic Spectroscopy}
\label{sec:systematics}

This study builds on the work described in \cite{DelDebbio:2008zf,DelDebbio:2010hu} where spectroscopic observables of MWT were measured through lattice
simulations. The computation was performed using the \emph{HiRep} code, designed to simulate theories of generic number of colours, and with fermions in a generic representation of the gauge group. The simulations used the Wilson gauge action, and the Wilson fermion formulation along with the RHMC algorithm. A number of lattice volumes have been analysed,
from $16\times8^3$ to $64\times24^3$ with a range of bare quark masses. The majority of the ensembles have been generated at $\beta=2.25$, although we do here present the results of some additional runs on the largest lattice at $\beta=2.1$.

For this study we have performed some alternative analyses to those in \cite{DelDebbio:2010hu}. The \emph{Chroma} suite of lattice software \cite{Edwards:2004sx} has been extended to operate with several fermionic representations other than the fundamental, including the adjoint. This will allow us to utilise the in-built smearing routines of \emph{Chroma} for our spectroscopic study.

\begin{figure}[!htp]
\centering
\subfloat[$D_1$]{
\includegraphics[scale=0.27]{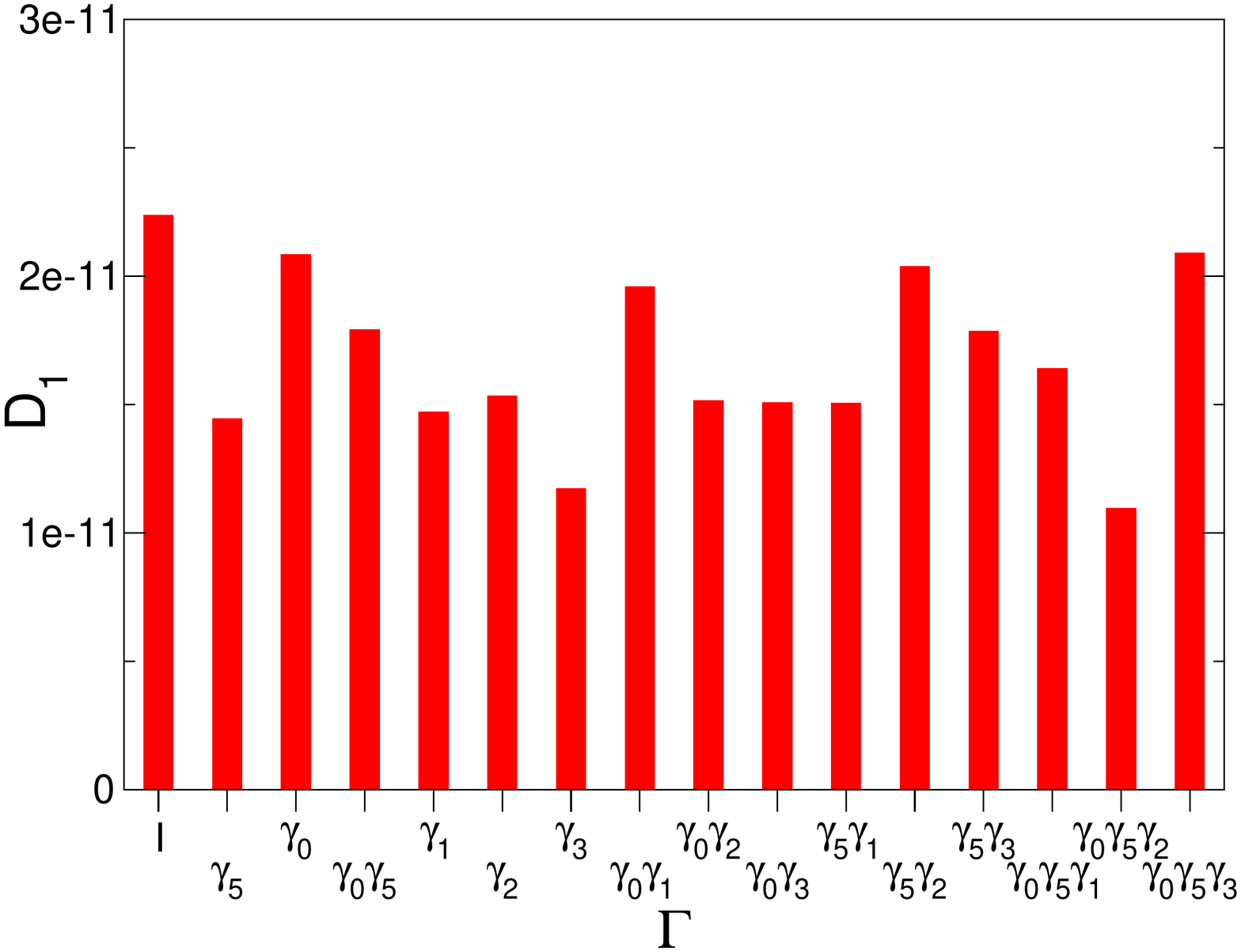}
}
\subfloat[$D_2$]{
\includegraphics[scale=0.27]{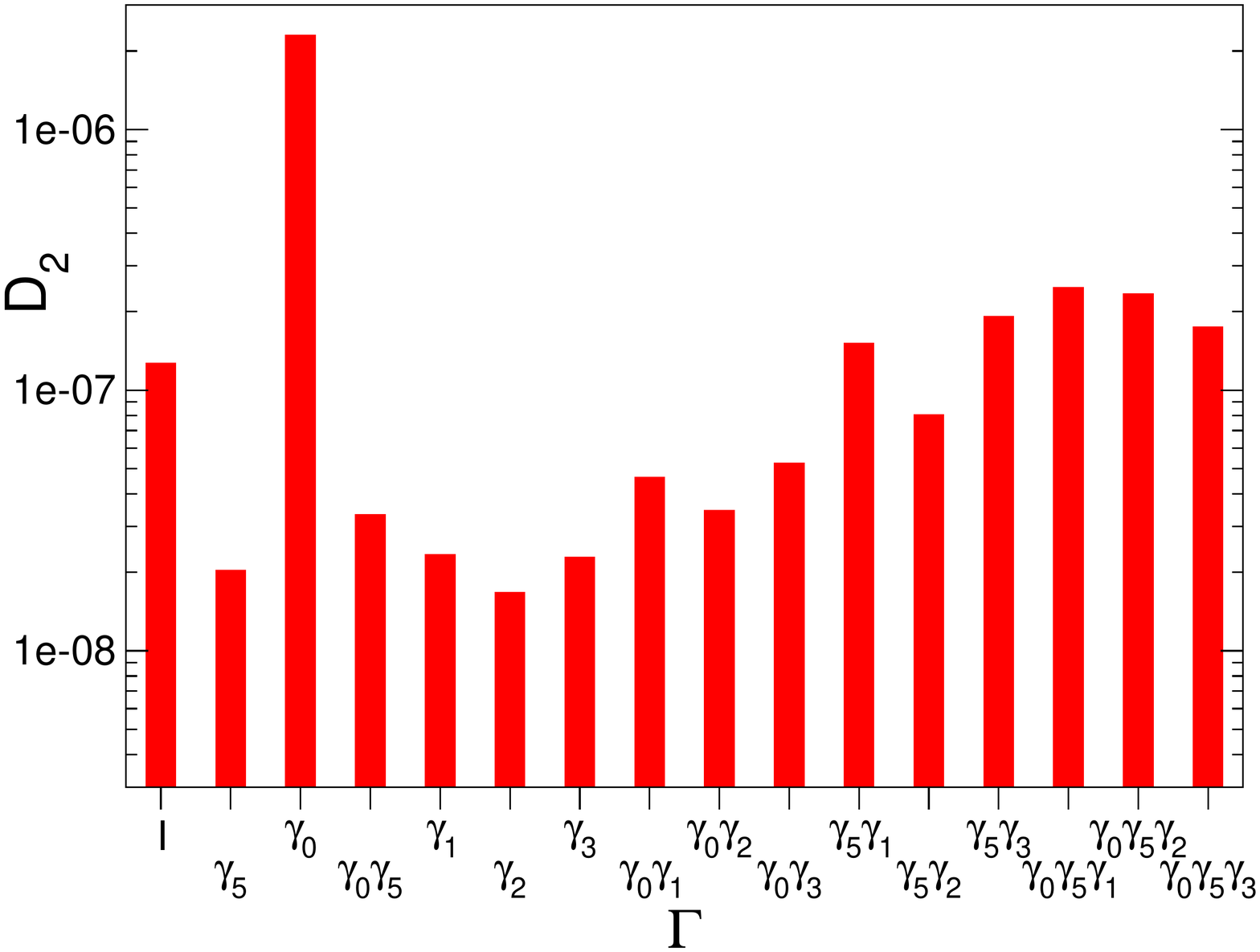}
}
\caption{Discrepancies between local correlators from \emph{HiRep} and \emph{Chroma}, computed to test the extension of \emph{Chroma} (for working with adjoint fermions) against HiRep.}
\label{discrepancies} 
\end{figure}

In order to test the modified \emph{Chroma}, we measured the local correlators as defined in (\ref{localcorr}), with $\Gamma=\Gamma^\prime$ both with 
\emph{HiRep} ($f^{(h)}_\Gamma(t)$) and \emph{Chroma} ($f^{(c)}_\Gamma(t)$). We used an ensemble of configurations on a $8\times4^3$ lattice with $\beta=2.25$
and fermion bare mass $am_0=-1$. Fig. \ref{discrepancies} illustrates the discrepancy between the two determinations, defined as:
\begin{align}
D_1=\sqrt{\sum_t\left(f^{(c)}_{\Gamma}(t)-f^{(h)}_{\Gamma}(t)\right)^2}&&
D_2=\sqrt{\sum_t\left(\frac{f^{(c)}_{\Gamma}(t)-f^{(h)}_{\Gamma}(t)}{f^{(h)}_\Gamma(t)}\right)^2}
\ .
\end{align}

We proceeded to utilise the in-built smearing routines found in \emph{Chroma} to perform measurements on the gauge configurations generated with \emph{HiRep} using a number of different quark smearings. We have investigated the use of both wall-smearing and a gauge invariant gaussian smearing, as defined in App. \ref{smearing}. Definitions of all observables discusses can be found in App. \ref{app:mesoncorr}.

Gaussian smearing involves two parameters, which can be chosen to optimize the technique. They are the width of the smearing function and the number of
applications of the smearing operator, which must be large enough to reasonably approximate the gaussian form. These two parameters have been adjusted in order
to maximize the overlap of the smeared operator with the ground state. On the other hand, the wall-smearing is a parameter-free procedure.

\begin{figure}[htp]
\centering
\subfloat[$16\times8^3$]{
\includegraphics[scale=0.27]{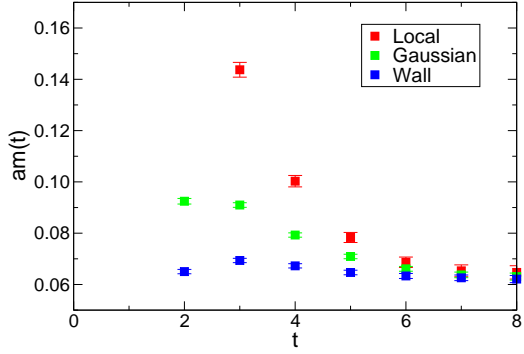}
\label{pcaccomp8}
}
\subfloat[$24\times12^3$]{
\includegraphics[scale=0.27]{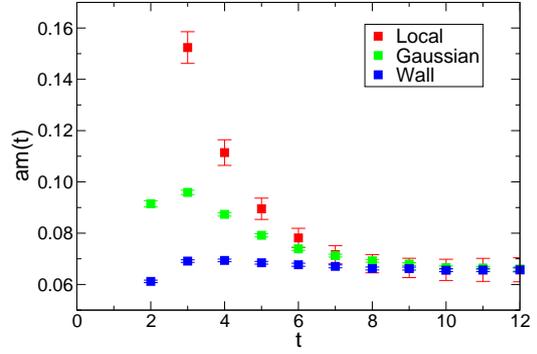}
\label{pcaccomp12}
}
\caption{Comparison of the PCAC mass from different smearings at $am_0=-1.175$.}
\label{pcaccomp}
\end{figure}

\begin{figure}[htp]
\centering
\subfloat[$16\times8^3$]{
\includegraphics[scale=0.27]{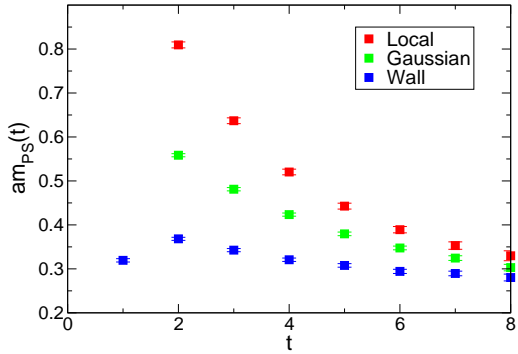}
\label{pscomp8}
}
\subfloat[$24\times12^3$]{
\includegraphics[scale=0.27]{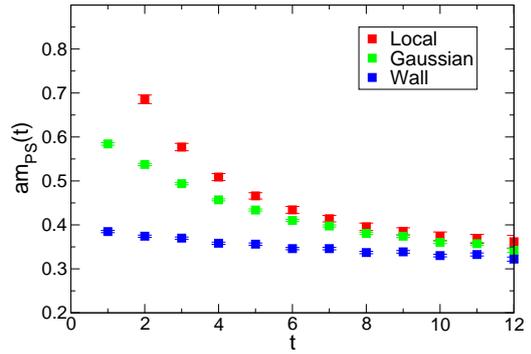}
\label{pscomp12}
}
\caption{Comparison of the pseudoscalar mass from different smearings at $am_0=-1.175$.}
\label{pscomp}
\end{figure}

\begin{figure}[htp]
\centering
\subfloat[$16\times8^3$]{
\includegraphics[scale=0.27]{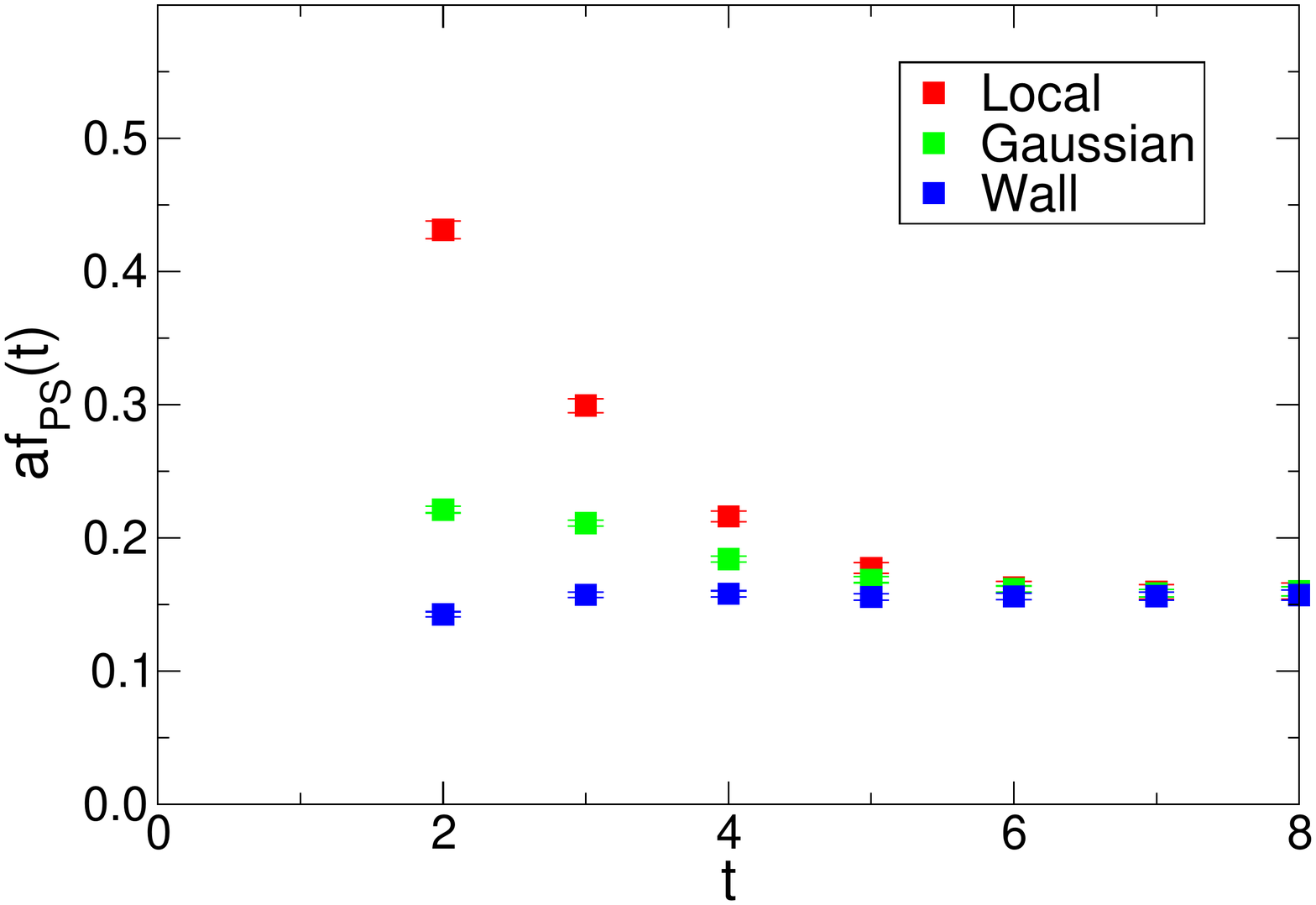}
\label{fpscomp8}
}
\subfloat[$24\times12^3$]{
\includegraphics[scale=0.27]{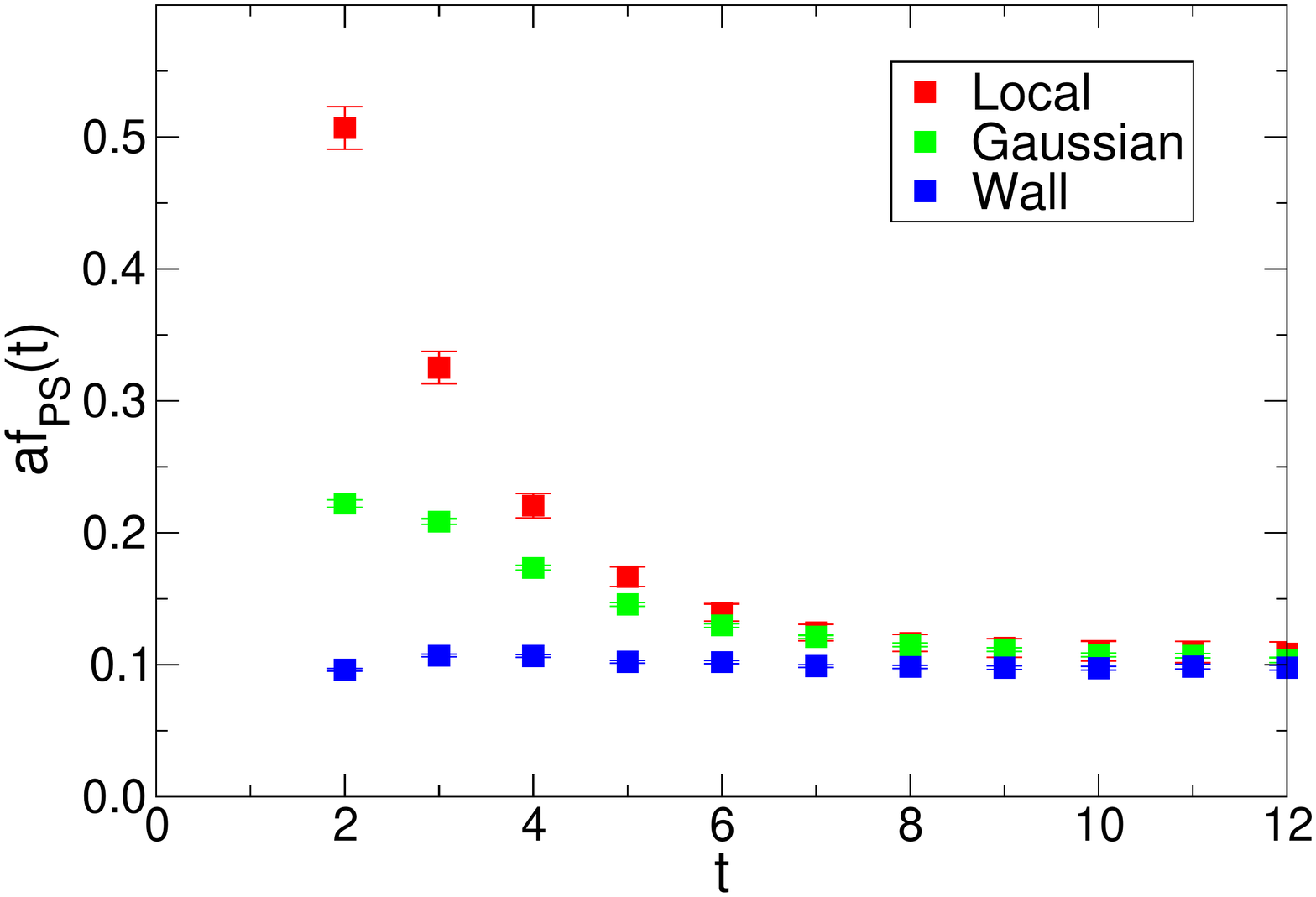}
\label{fpscomp12}
}
\caption{Comparison of the pseudoscalar decay constant from different smearings at $am_0=-1.175$.}
\label{fpscomp}
\end{figure}

We systematically compared local, gaussian (with optimised parameters) and wall-smeared sources on our ensembles. At our lightest masses, the wall-smeared
sources have the largest overlap with the ground state, which is reflected in the flattest effective masses. In Figs.~\ref{pcaccomp}, \ref{pscomp},
\ref{fpscomp} we show respectively the PCAC and PS effective masses and the PS effective decay constant computed with the three methods.

Since we are mainly interested in the light masses, we will focus on the wall-smeared results in the rest of this work.

\section{Effectiveness of the wall-smeared sources}
\label{sec:evaluate}

Using smeared sources allows us to choose an operator with a larger projection onto the ground state of a given channel.  The wave-function of the ground state is spread over many lattice sites, and we can improve the overlap of the operator with the ground state by giving a spatial size to the source. The smeared correlator will be less contaminated by the excited states, and therefore it will be characterized by a single $\cosh$ signal for a larger temporal separation than the one constructed with point operators. This is reflected in a longer plateau in the effective mass.
On the contrary one drawback of using smeared sources is that it makes the analysis more sensitive to the algorithm's autocorrelation time.
In this section we propose a quantitative study of these two aspects: the behaviour of the size of the plateaux for different kinds of sources, and the autocorrelation time connected with the use of these sources.

\subsection{Autocorrelations}

Correlators generated using sources with an extended spatial profile are expected to be associated with longer autocorrelation times, due to the fact that the low energy modes of the fields need more Monte-Carlo time to propagate. This effect is observed throughout our study, indeed the autocorrelation time associated with the results from smeared correlators is generically at least of the order of twice that of those involved with the local correlators. This is supported both by the direct measurement of the integrated autocorrelation time~\cite{Madras:1988ei} associated with the observables, and by the analysis of the behaviour of the standard deviation of the observables.

Both the aforementioned studies have been performed by grouping the $N$ data into $N/b$ blocks of a given length $b$. A reduced dataset of length $N/b$ is created by averaging the required statistic over each block. A bootstrap analysis is then performed on the reduced dataset. By increasing the block size $b$, we are creating effective estimates less and less autocorrelated, hence when the block size is bigger than the autocorrelation we expect to see a plateau appearing in the standard deviation, signaling that the reduced dataset is decorrelated. We observe that the plateau starts at a block size corresponding to an integrated autocorrelation time of order 1.

\begin{figure}[!htp]
\centering
\includegraphics[scale=0.3]{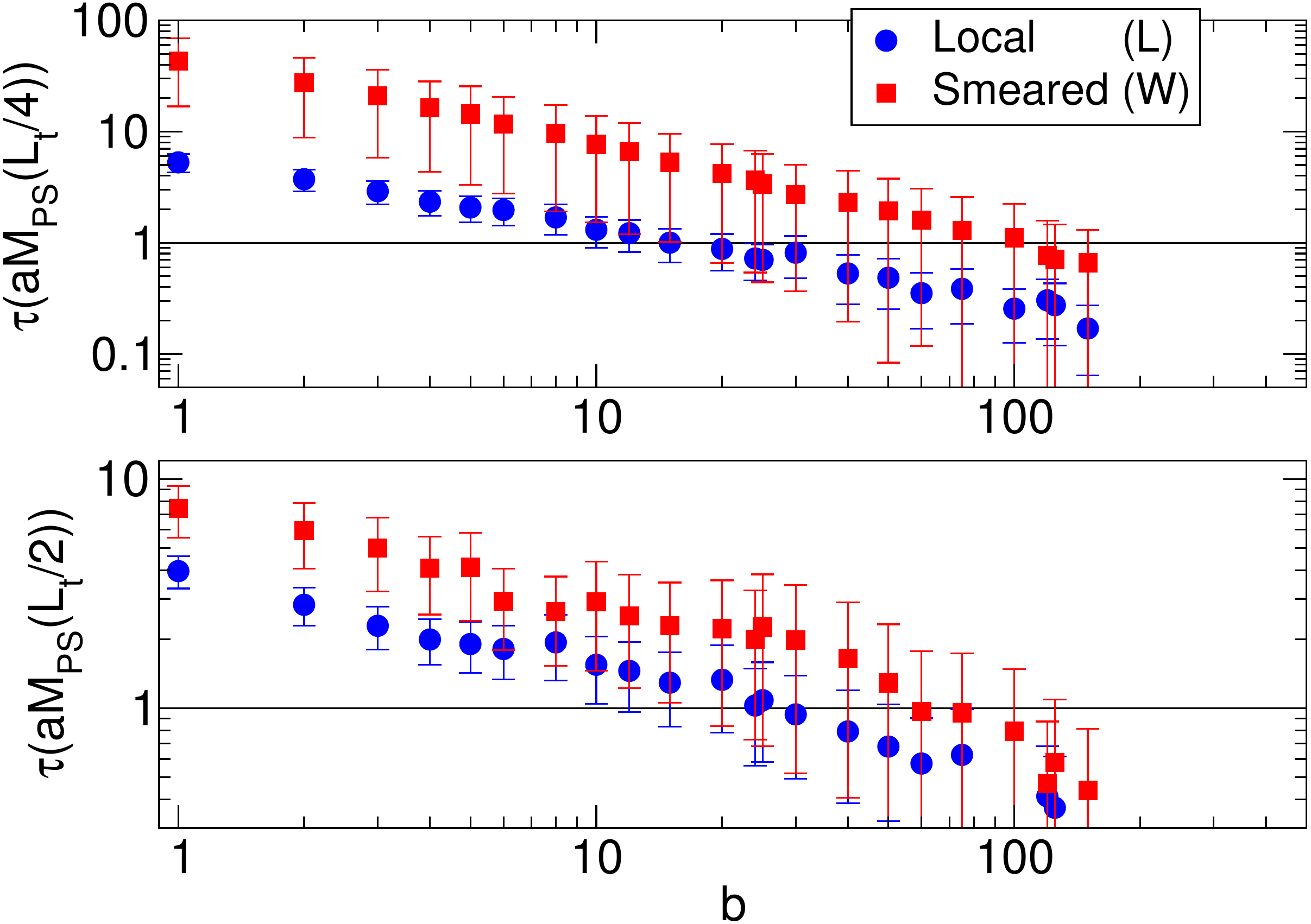}
\hspace{.8cm}
\includegraphics[scale=0.3]{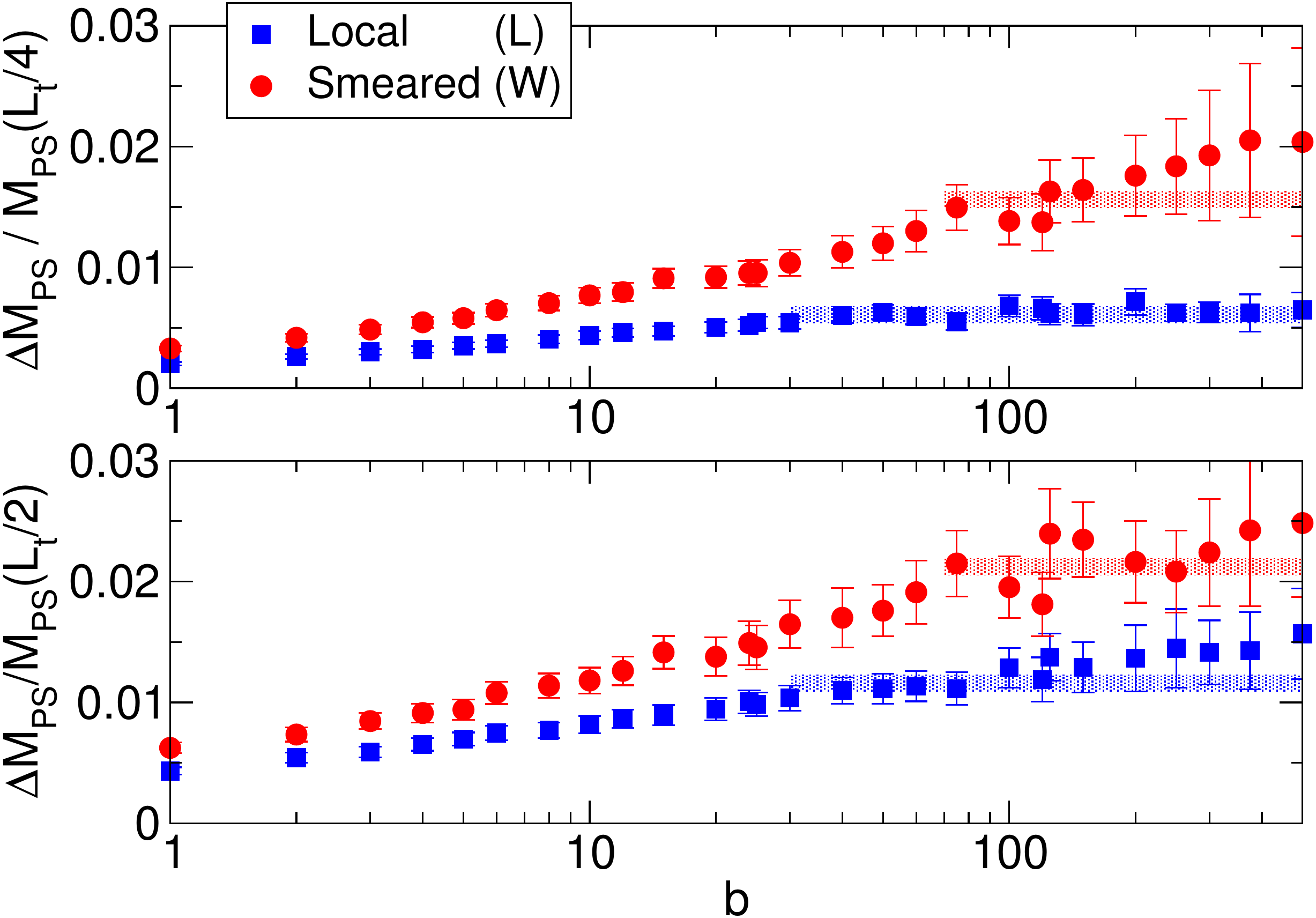}
\caption{Autocorrelation analysis conducted on a $24\times12^3$ lattice at $am_0=-1.175$, for the PS effective mass in two temporal points. In the left panel, integrated autocorrelation time as a function of the block size $b$. In the right panel, relative error as a function of the block size $b$. The plateaux of the relative error are highlighted with shadowed rectangles. The plateaux in the relative error set in when the integrated autocorrelation time becomes of order 1.
}
\label{autocorrplots}
\end{figure}

Our analysis of the autocorrelation is illustrated in Fig.~\ref{autocorrplots}, for the PS effective mass obtained with both local (L), and wall-smeared (W) sources, evaluated at two temporal points.

From the left panel of Fig.~\ref{autocorrplots} we see that the measured autocorrelation time for the smeared results are generically larger than those for the local results. From the right panel of Fig.~\ref{autocorrplots} we see that the standard deviation of our observable increases for both sets of correlators as we increase the block size from zero, up to a point where it appears to reach a plateau for a significant range of $b$ for both cases. The value of $b$ where this plateau sets in is interpreted as the length in simulation time over which the data are uncorrelated. From the right panel of Fig.~\ref{autocorrplots} we would conclude that the autocorrelation time of our local result was $\sim30$ while that of the smeared result was $\sim80$. Indeed returning to the left panel of Fig.~\ref{autocorrplots} we see that at this value of $b$, the corresponding value of the integrated autocorrelation time is close to 1, which supports our conclusion.

This picture is replicated across our ensembles, and we have accounted for this in our results by conducting our bootstrap analysis over appropriately reduced datasets.

\subsection{Plateaux of the effective masses}
\label{sec:plateaux}

If the smearing procedure is effectively suppressing the contribution of the excited states to the correlators, one has to observe the effective masses flattening around the midpoint $t=L_t/2$, and the plateaux becoming longer when visible. We can quantitatively estimate the flatness of the effective mass using the absolute value of the incremental ratio of the effective mass between $t=L_t/2$ and $t=L_t/2-\Delta t$:
\begin{equation}
\frac{\Delta m_{PS}}{\Delta t} \equiv \left| \frac{m_{PS}(L_t/2-\Delta t) - m_{PS}(L_t/2)}{\Delta t} \right| \ .
\end{equation}
A value for $ \Delta m_{PS}/\Delta t $ compatible with zero implies that the plateau in the effective mass is long at least $\Delta t$ points. For very small values of $\Delta t$ the incremental ratio is dominated by the statistical error. On the other hand the effective masses obtained with smeared sources are sometimes non monotonic. In this case the incremental ratio defined with a too large value for $\Delta t$ is not a good estimate for the flatness of the plateau. An intermediate range of values for $\Delta t$ exists, in which our analysis makes sense. We explicitly checked that our conclusions do not change choosing $\Delta t$ in such a range, and we chose $\Delta t=4$ for definiteness.

In general the smaller $ \Delta m_{PS}/\Delta t $ is, the flatter the plateau. Notice that it is important to take the absolute value in the definition above: while the effective mass defined from local correlators is always decreasing, it is not so for smeared correlators.

\begin{figure}[!htp]
\centering
\subfloat[$16\times8^3$]{
\includegraphics[scale=0.27]{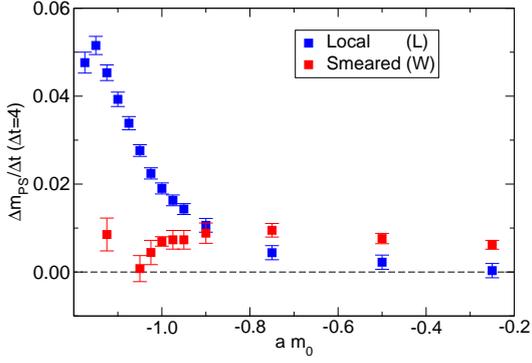}
\label{eval8}
}
\subfloat[$24\times12^3$]{
\includegraphics[scale=0.27]{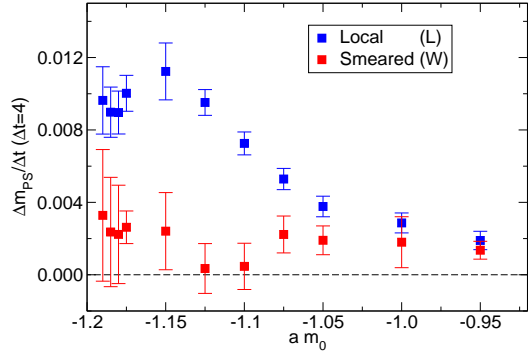}
\label{eval12}
}

\subfloat[$32\times16^3$]{
\includegraphics[scale=0.27]{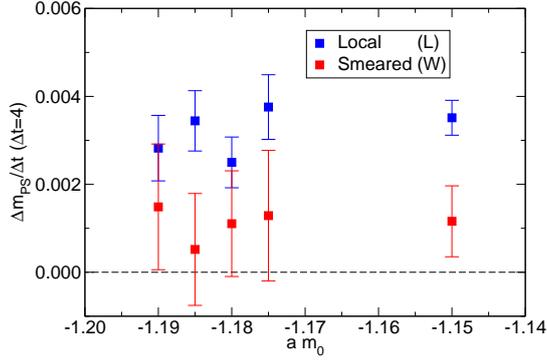}
\label{eval16}
}
\caption{Incremental ratio $ \Delta m_{PS}/\Delta t $ as a function of the bare mass. The smaller this quantity, the better the quality of the plateau of the PS effective mass. On the $16 \times 8^3$ lattice, the local correlators give flatter plateaux for bare masses larger than $-0.8$, while the smearing is effective below $-0.9$. On the $24 \times 12^3$ lattice, the local and smeared sources give plateaux of similar quality for the two heaviest masses, while the smearing is effective for all the other masses. Finally the smearing is always effective on the $32 \times 16^3$.}
\label{plateval}
\end{figure}

In Fig.~\ref{plateval}, the quantity $ \Delta m_{PS}/\Delta t $ is plotted for all our pseudoscalar effective masses on the $16\times8^3$, $24\times12^3$ and
$32\times16^3$ lattices, both for local and wall-smeared correlators.

One expects that at small masses the wave function of the pseudoscalar meson is more spread, hence the wall-smeared source should have a larger overlap with the ground state. On the contrary at large masses the wave function is more localised therefore the local sources should work better. Our analysis presented in Fig.~\ref{plateval} substantiates this expectation. On the $16\times8^3$ lattice the wall-smeared sources give better or comparable plateaux than the local sources for masses $am_0 \le -0.9$. On the $24\times12^3$ and $32\times16^3$ lattices the wall-smeared sources are to be (sometimes marginally) preferred to the local ones for all the simulated masses.

In the presentation of the results obtained from wall-smeared sources we will always cut the masses in the $16\times8^3$ lattice for which the local sources are
actually preferable to the wall-smeared ones, unless otherwise stated.

Finally, we point out that the same analyses using the effective V meson mass and the effective PS decay constant produce very similar results.

\section{Results}
\label{sec:results}

In the present section, we will present our results for the mesonic observables from the wall-smeared sources. Complete results of all observables analysed are
also presented in Sec. \ref{app:results}. Although only the results at $\beta=2.25$ will be discussed in detail, measurements at
$\beta=2.1$ can also be found in the tables. For the full local results, the reader is referred to~\cite{DelDebbio:2010hu}.

We will consider only those fermionic masses for which the wall-smeared sources give an improvement on the plateaux of the effective masses with respect to the
local sources, as discussed in Sec.~\ref{sec:plateaux}. For all these masses, the wall-smeared results have to be trusted more than the local ones. The
disagreement between the two determinations gives an estimate of the systematic error due to a bad determination of the plateaux, mainly affecting our previous
results obtained from the local sources.

In order to quantify this disagreement we use two different estimators: the \textit{pull} and the \textit{relative discrepancy}. We will denote $O_L \pm
\Delta O_L$ and $O_S \pm \Delta O_S$ the determination of the generic observable $O$ using respectively local and smeared sources. The pull estimates the
relative size of the systematic and statistical errors and is defined as:
\begin{equation}
P(O) = \frac{| O_L - O_S |}{\sqrt{\Delta O_L^2 + \Delta O_S^2}} \ .
\label{eq:pull}
\end{equation}
A small value for the pull is desirable, indicating that the systematic errors are smaller than the statistical ones. However a small value for the pull can be
obtained either with a small systematic error or with a large statistical one. Therefore it is not an absolute estimator of the goodness of a measurement. The
relative discrepancy estimates the systematic error, relative to the average of the two determinations:
\begin{equation}
D(O) = \frac{2 | O_L - O_S |}{O_L + O_S} \ .
\label{eq:discr}
\end{equation}
A small value for the relative discrepancy indicates that the systematic effects contribute to a small fraction of the determination of the observable $O$.

In what follows, we will consider separately the PCAC quark mass, the PS and V masses and their ratios, the PS and V decay constant. Again, we refer the reader to
Appendix~\ref{app:mesoncorr} for the definition of these observables. We will present the results for the wall-smearing sources, and we will discuss the
differences with the local-source results using the pull and the relative discrepancy.

\subsection{PCAC mass}

\begin{figure}[!htp]
\centering
\includegraphics[scale=0.38]{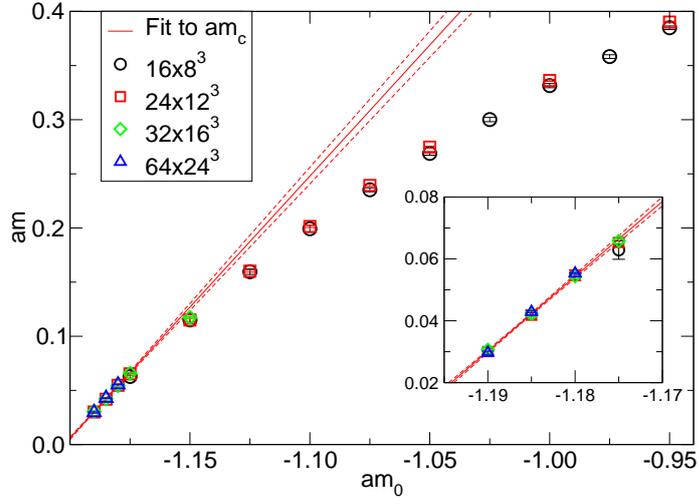}
\caption{PCAC quark mass for ensembles at $\beta=2.25$, computed with wall-smeared sources, as a function of the quark bare mass. The result of the linear fit for extracting the critical bare mass is also shown. In the inset, the lightest masses are zoomed in.}
\label{pcacres}
\end{figure}

In Fig. \ref{pcacres} results for the PCAC mass from the wall-smeared correlators on all $\beta=2.25$ ensembles are presented. The inset illustrates a close up of the approach to the chiral limit, with a linear extrapolation to zero quark mass. Using this we find the critical bare quark mass to be $am_c=-1.2022(14)$, from a fit using the three lightest points on the $24\times12^3$ lattice, which compares very well to the result obtained from the local data \cite{DelDebbio:2010hu}.

\begin{figure}[!htp]
\centering
\includegraphics[scale=0.38]{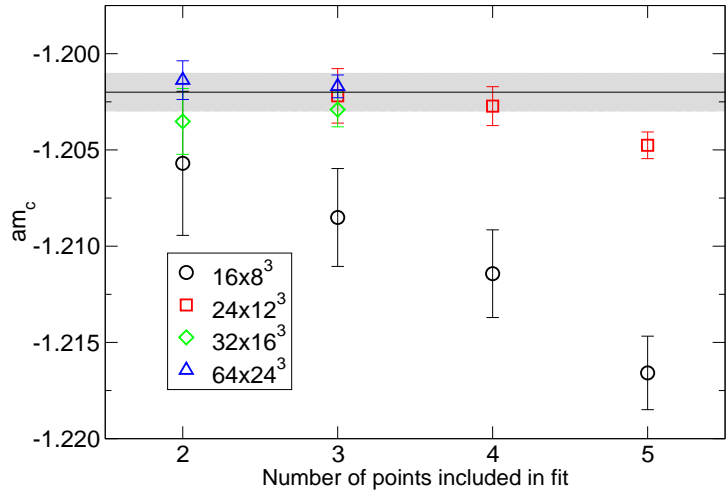}
\caption{Critical quark mass extracted from a linear fit with different fit ranges. The critical mass as obtained from local data is shown as a grey band.}
\label{mcplot}
\end{figure}

In Fig.~\ref{mcplot} we show the stability of this fit against varying the number of points used. We compare this to the result obtained from local correlators,
noting the agreement. It is also clear that finite volume effects for this quantity are at most comparable with the statistical uncertainty. 

\begin{figure}[!htp]
\centering
\subfloat{
\includegraphics[scale=0.27]{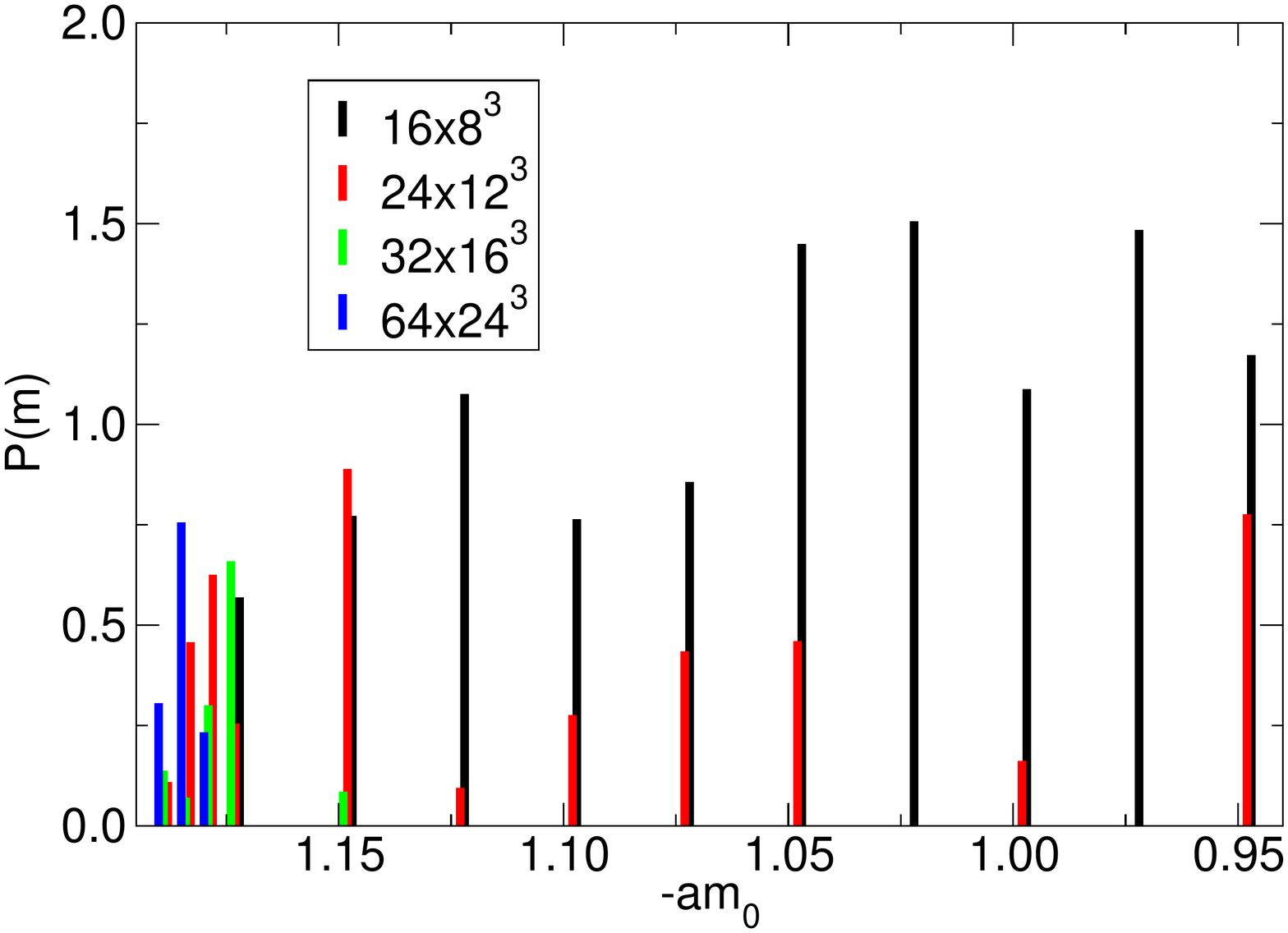}
}
\subfloat{
\includegraphics[scale=0.27]{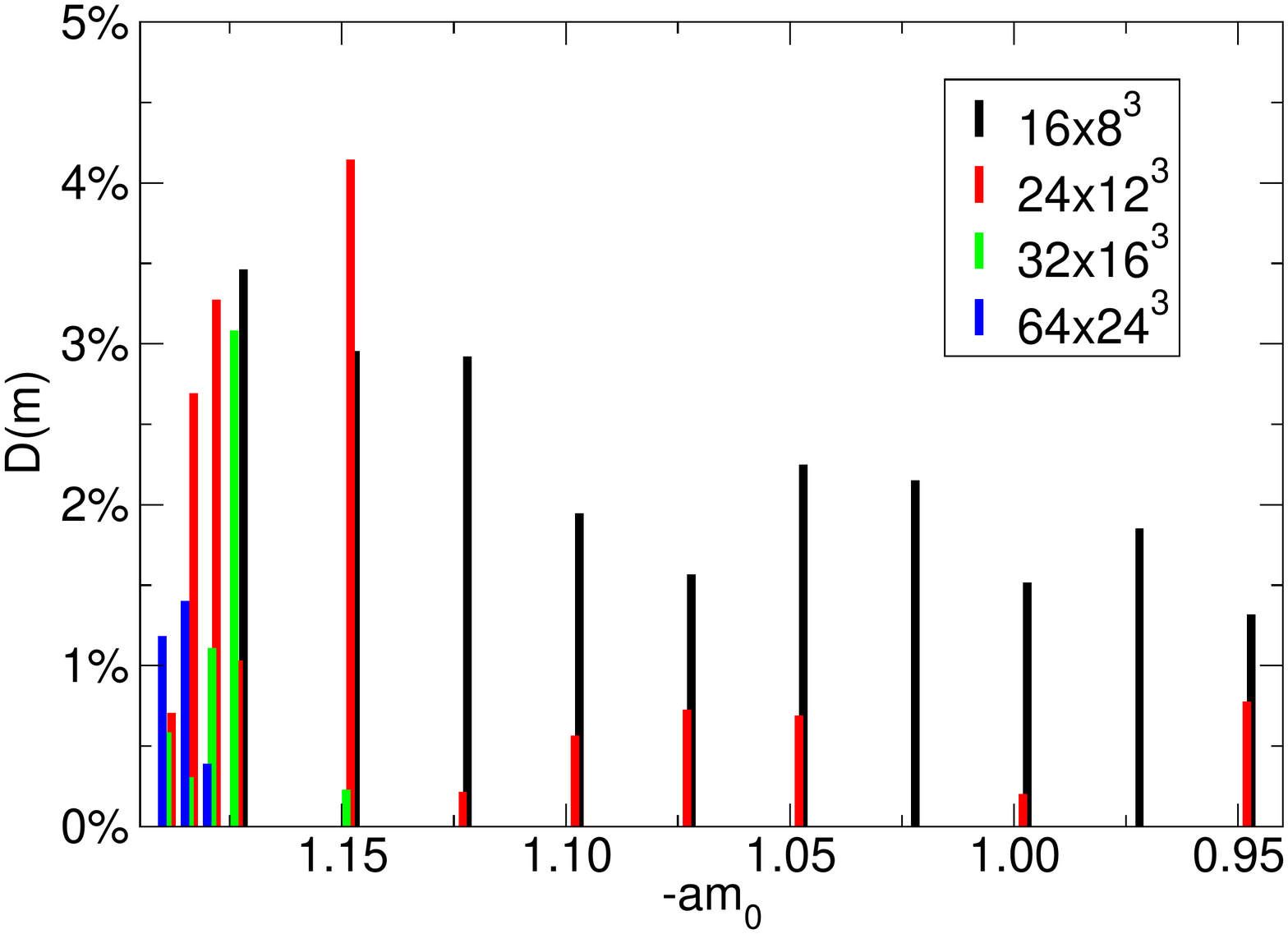}
}
\caption{Pull and relative discrepancy as defined in Eqs.~\eqref{eq:pull} and~\eqref{eq:discr} for the PCAC quark mass  ($\beta=2.25$).}
\label{fig:pullmpcac}
\end{figure}

In Fig.~\ref{fig:pullmpcac} we show the pull and the relative discrepancy as defined in Eqs.~\eqref{eq:pull} and~\eqref{eq:discr} between the local and wall-smeared determinations of the PCAC quark mass. We include all the masses at which the wall-smeared sources give an improvement of the plateaux in the effective masses over the local sources. As shown in the left panel of Fig.~\ref{fig:pullmpcac}, the pull is always smaller than 1 (or marginally larger than 1 for the smallest volume), indicating that the systematic error due to a short temporal direction is of the order of the statistical uncertainty. The right panel of Fig.~\ref{fig:pullmpcac} shows that the systematic error is of order of a few percents for the PCAC mass.

\subsection{Meson masses}

\begin{figure}[!htp]
\centering
\includegraphics[scale=0.38]{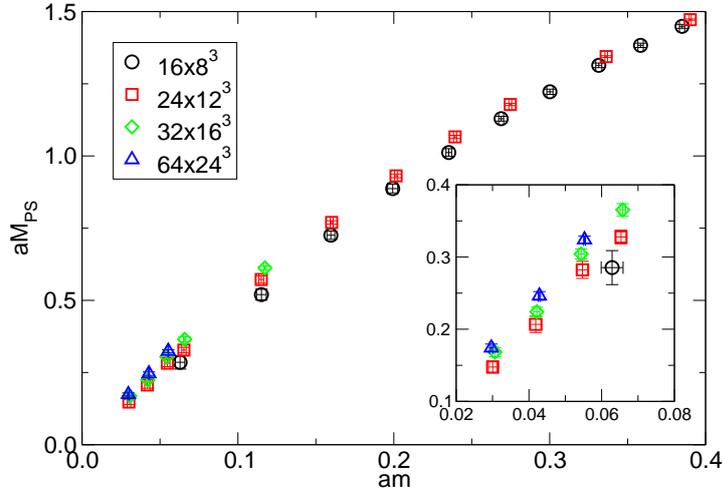}
\caption{Pseudoscalar meson mass for ensembles at $\beta=2.25$, computed with wall-smeared sources, as a function of the PCAC mass. In the inset, the lightest masses are zoomed in.}
\label{mpsplot}
\end{figure}

\begin{figure}[!htp]
\centering
\includegraphics[scale=0.38]{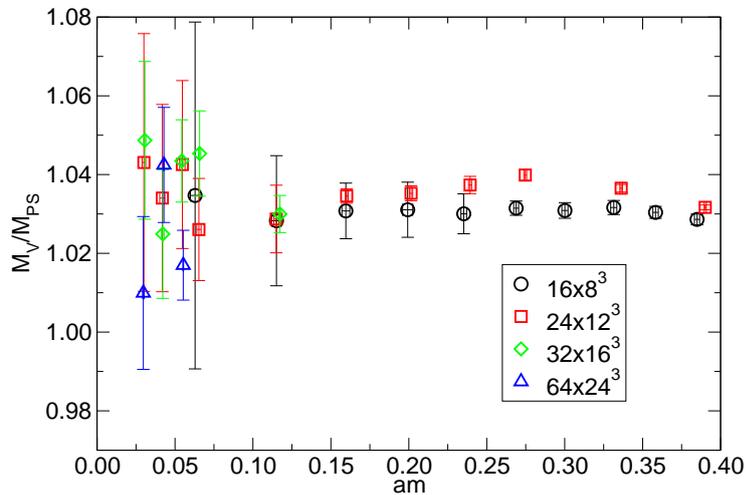}
\caption{Ratio of $M_{V}$ to $M_{PS}$ for ensembles at $\beta=2.25$, computed with wall-smeared sources, as a function of the PCAC mass. The plateau in this ratio at small masses has been interpreted in our previous works~\cite{DelDebbio:2009fd,DelDebbio:2010hu} as a signal for IR-conformality. The smeared sources have amplified the finite volume effects at masses around $am \simeq 0.3$. This effect will be discussed in Sec.~\ref{sec:finitevolume}}
\label{mvmps_plot}
\end{figure}

Fig. \ref{mpsplot} shows the results obtained for the pseudoscalar mass $M_{PS}$ as a function of the PCAC quark mass $m$, from the $\beta=2.25$ data.
Fig.~\ref{mvmps_plot} shows the ratio $M_V/M_{PS}$. We recall that the existence of a plateau at small masses of this ratio was one of the main ingredients for
arguing in favour of an IR fixed point in~\cite{DelDebbio:2009fd} and~\cite{DelDebbio:2010hu}. We notice that the smeared results stabilize the plateaux at very
small masses (especially by smoothing the behaviour of the largest volumes), while making more visible some finite-volume effects at intermediate masses. We
will discuss the finite-volume effects in Sec.~\ref{sec:finitevolume}.

\begin{figure}[!htp]
\centering
\subfloat{
\includegraphics[scale=0.27]{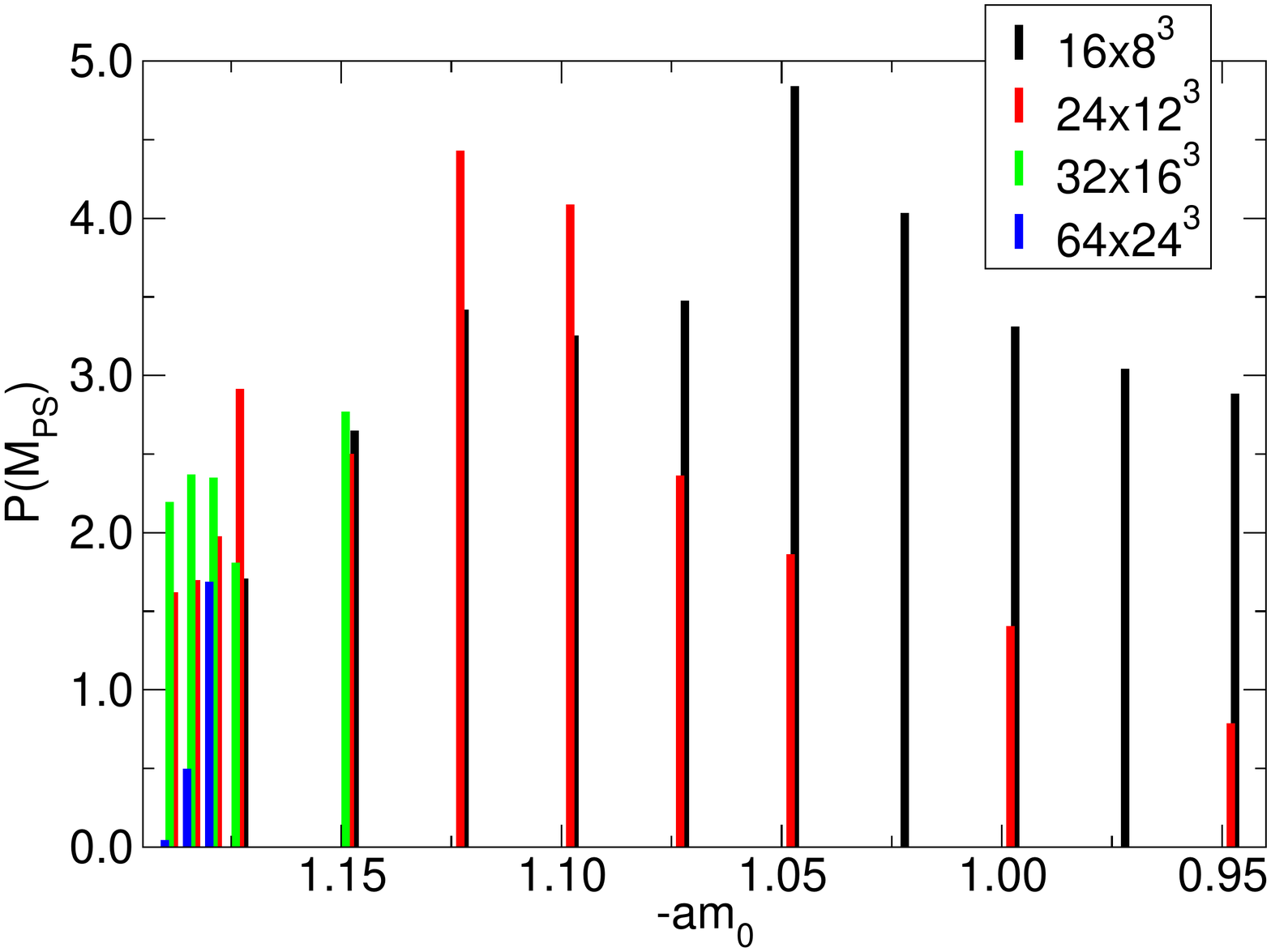}
}
\subfloat{
\includegraphics[scale=0.27]{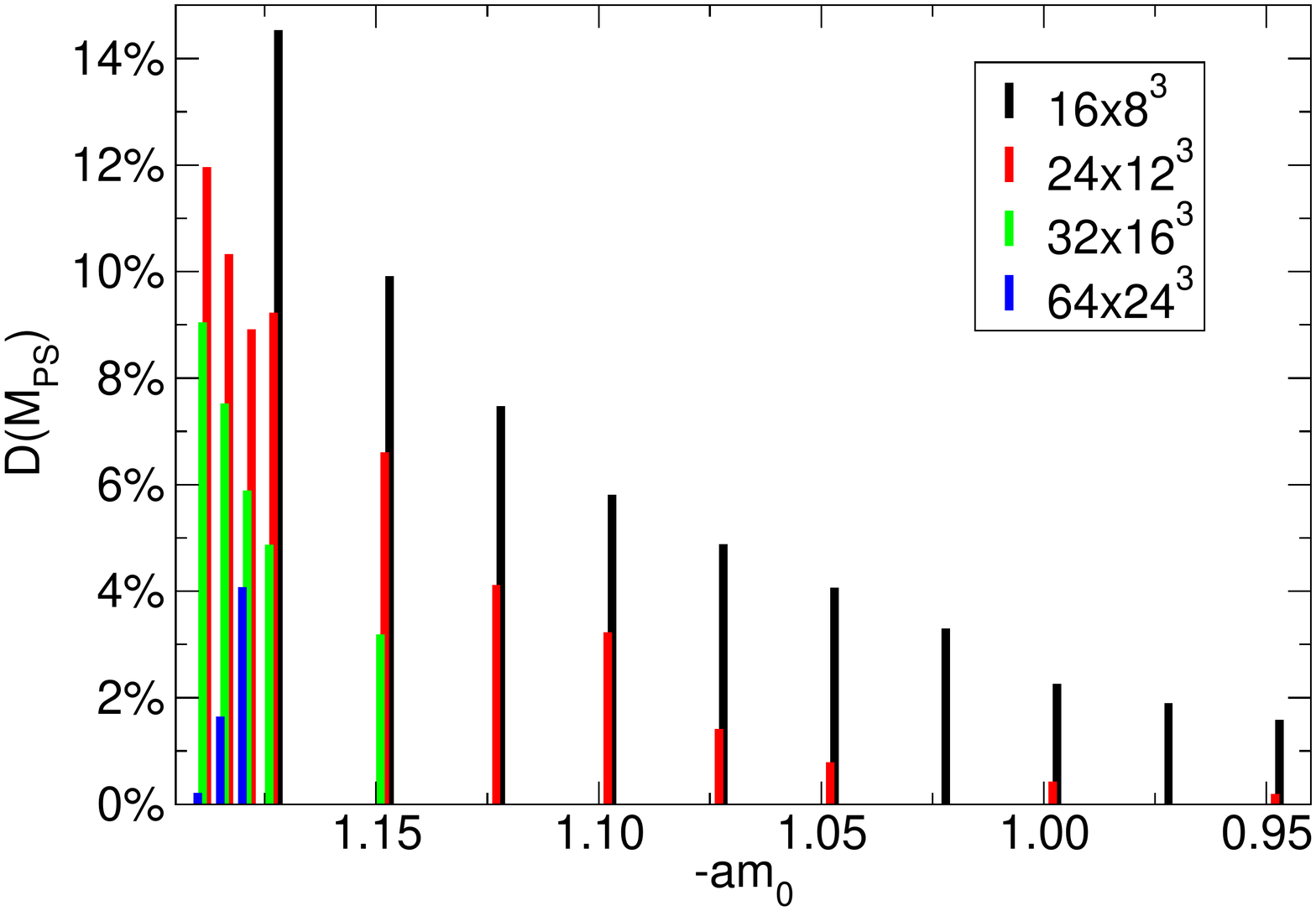}
}
\caption{Pull and relative discrepancy as defined in Eqs.~\eqref{eq:pull} and~\eqref{eq:discr} for the PS mass  ($\beta=2.25$).}
\label{fig:pullmps}
\end{figure}

We also report the pull and relative discrepancy as defined in Eqs.~\eqref{eq:pull} and~\eqref{eq:discr} between the local and wall-smeared determinations of
the PS mass in Fig.~\ref{fig:pullmps}. Again, we include all the masses at which the wall-smeared sources give an improvement of the plateaux in the effective
masses over the local sources. The local and smeared sources give quite different results at small masses. The relative discrepancy has a very regular
behaviour: it is larger for lighter masses or smaller volumes. For bare masses below $-1.15$ one has to use lattices larger than the $24 \times 12^3$ in order
to keep the relative discrepancy below the $10\%$ level. Although the relative discrepancy can get fairly large at these masses, the pull is always below $3$
which means that the two determinations are compatible within the $3\sigma$ range. This effect is generated by an increase of the relative statistical error at
light masses.

\begin{figure}[!htp]
\centering
\subfloat{
\includegraphics[scale=0.27]{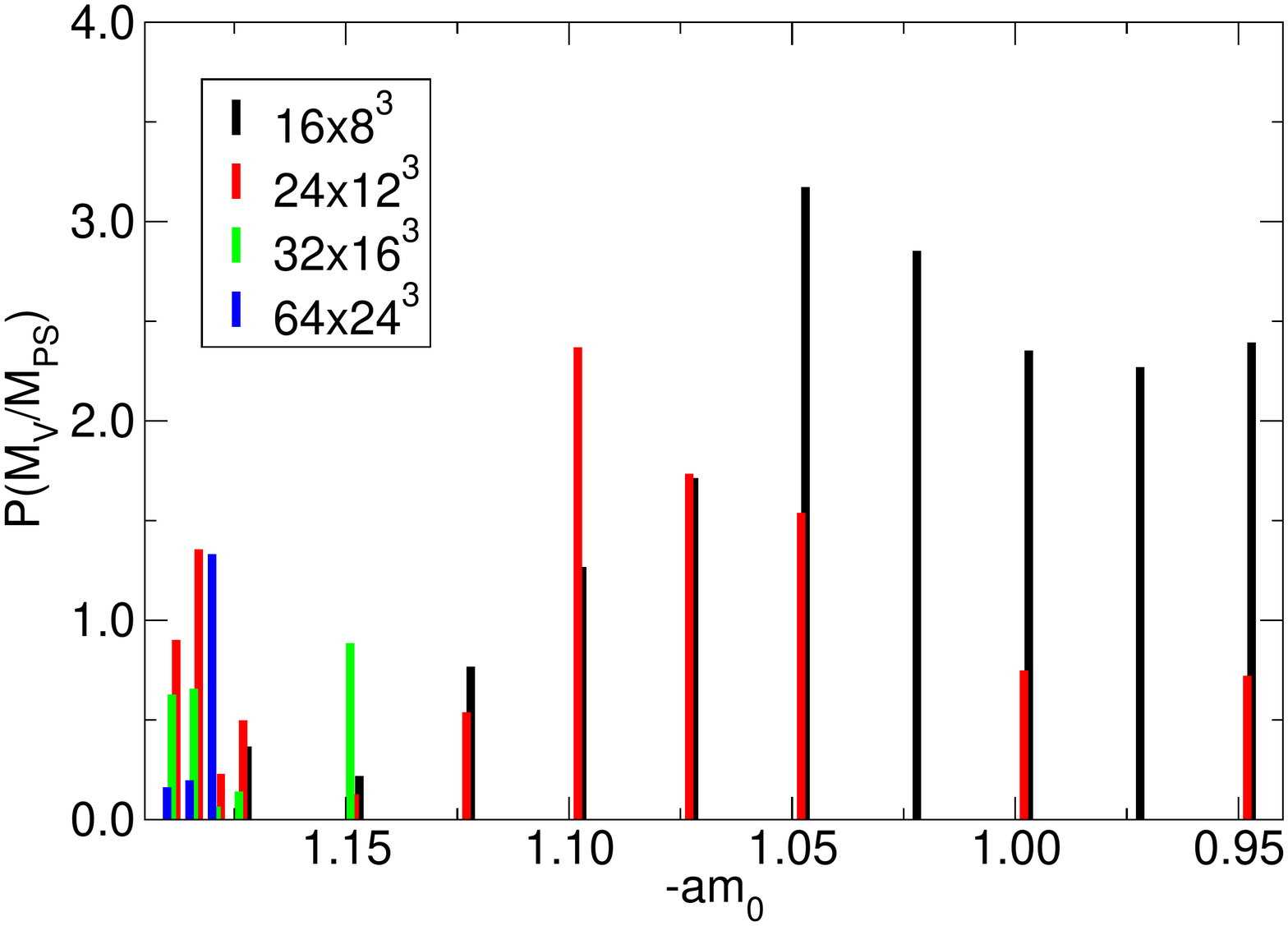}
}
\subfloat{
\includegraphics[scale=0.27]{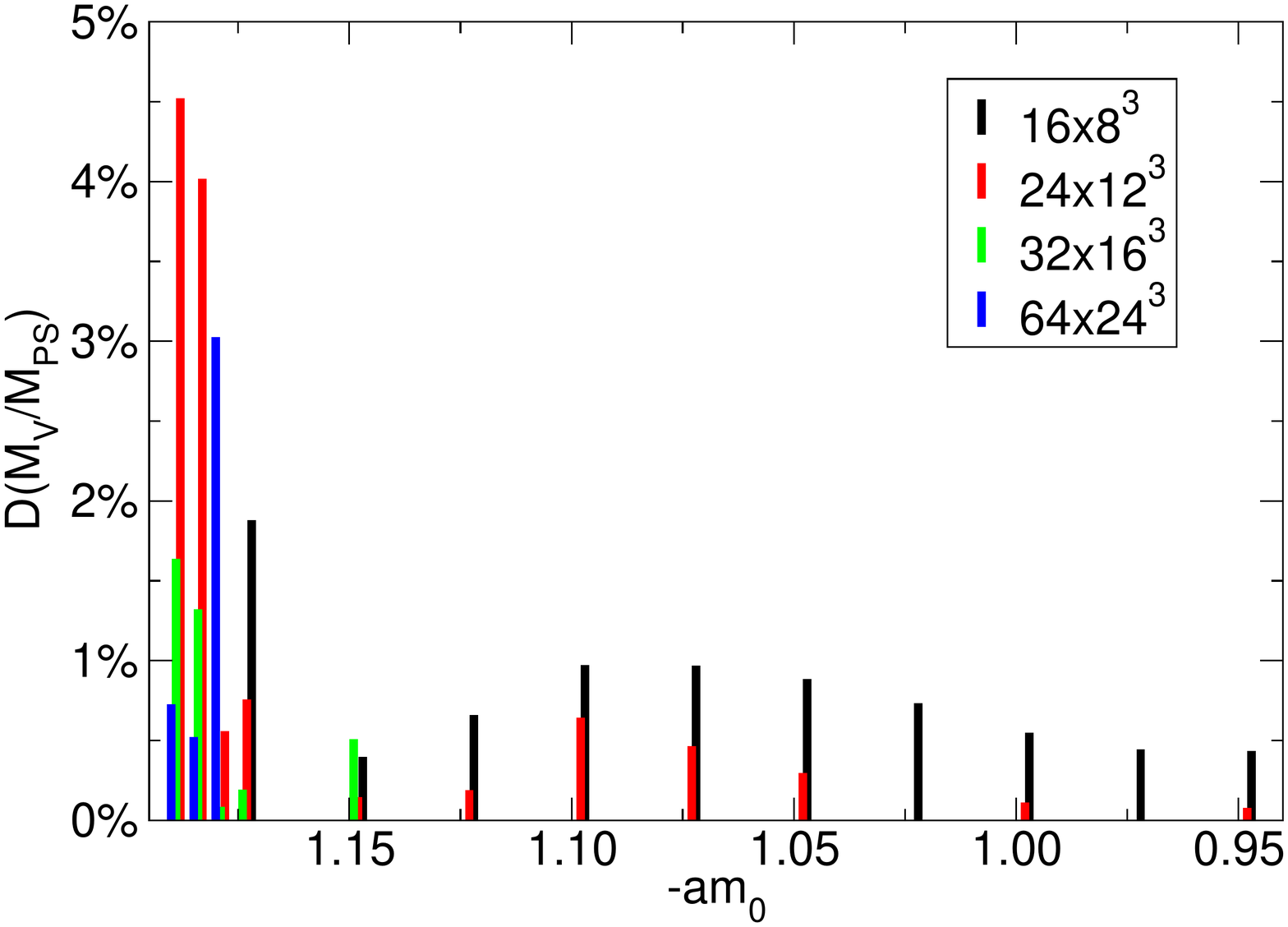}
}
\caption{Pull and relative discrepancy as defined in Eqs.~\eqref{eq:pull} and~\eqref{eq:discr} for the ratio of the V mass over the PS mass  ($\beta=2.25$).}
\label{fig:pullmvmps}
\end{figure}

Fig.~\ref{fig:pullmvmps} shows the pull and relative discrepancy between the local and wall-smeared determinations of the $M_V/M_{PS}$ ratio. The situation is
better here. The central values of the two determinations never differ by more than $5\%$ (relative discrepancy), and they are generally compatible (except
the smallest volume) within the $2\sigma$ range (pull).

\subsection{Decay constants}

\begin{figure}[!htp]
\centering
\subfloat{
\includegraphics[scale=0.27]{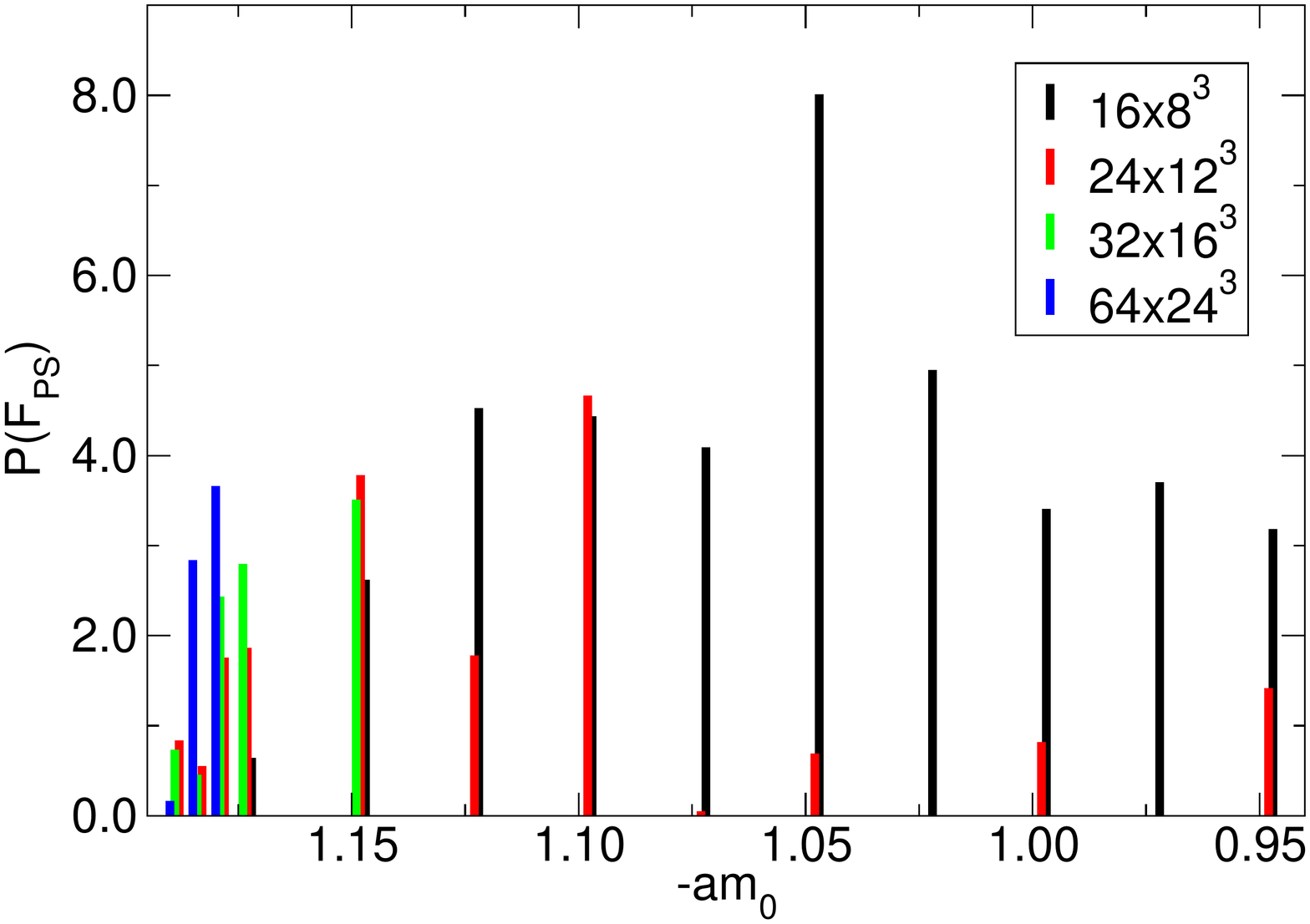}
}
\subfloat{
\includegraphics[scale=0.27]{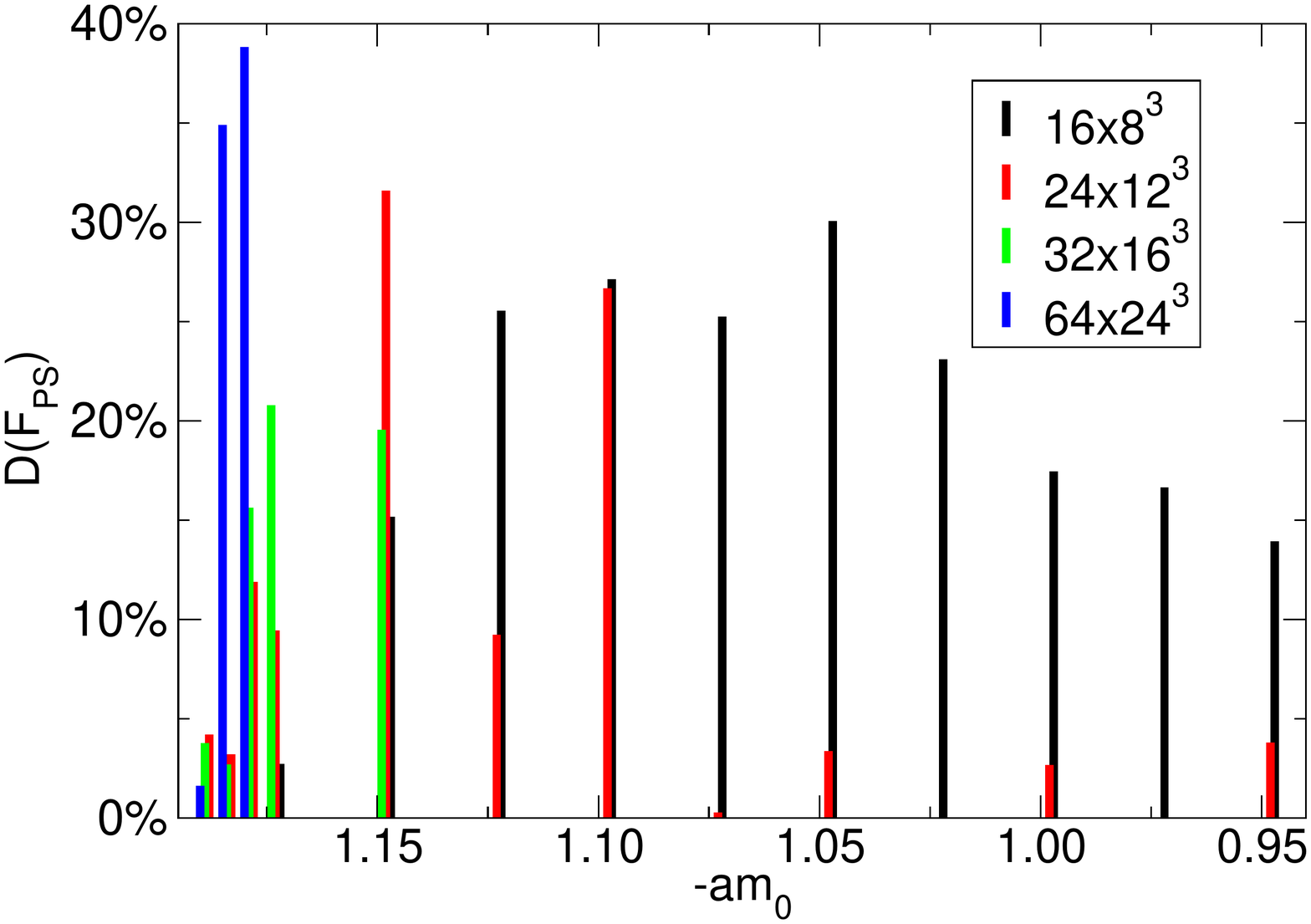}
}
\caption{Pull and relative discrepancy as defined in Eqs.~\eqref{eq:pull} and~\eqref{eq:discr} for the PS decay constant  ($\beta=2.25$).}
\label{fig:pullfps}
\end{figure}

Among the observables considered in this study, the PS decay constant is the quantity most affected by systematic errors due to a short temporal dimension. The
relative discrepancy between the local and smeared determinations (Fig.~\ref{fig:pullfps}) is almost always very large. On the $24\times 12^3$, $32\times 16^3$
and $64\times 24^3$ lattices, this large relative discrepancy is partly compensated by a large statistical error. In most of the cases the two determinations
are compatible (sometimes marginally) within $3\sigma$ of the statistical uncertainty (pull). On the $16\times 8^3$ lattice, the difference is more dramatic.
However for intermediate masses, the wall-smeared source gives a better defined plateau in the effective PS decay constant as discussed in
Sec.~\ref{sec:plateaux}, and therefore the smeared results have to be trusted more than the local ones.

\begin{figure}[!htp]
\centering
\includegraphics[scale=0.38]{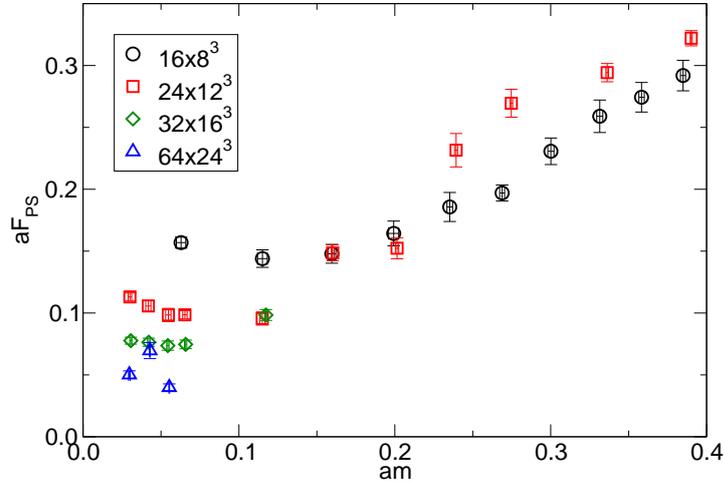}
\caption{Pseudoscalar decay constant for ensembles at $\beta=2.25$, computed with wall-smeared sources, as a function of the PCAC mass.}
\label{fpsplot}
\end{figure}

\begin{figure}[!htp]
\centering
\includegraphics[scale=0.38]{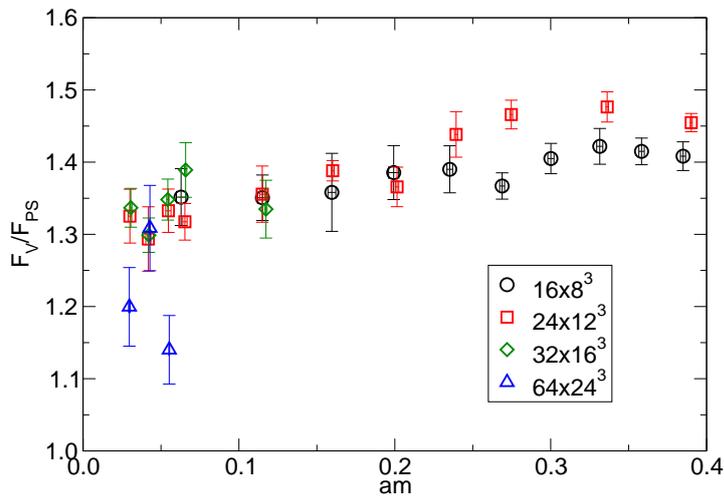}
\caption{Ratio of vector and pseudoscalar decay constants.}
\label{fvfpsplot}
\end{figure}

Fig.~\ref{fpsplot} shows the results for the PS decay constant from wall-smeared sources. The difference between the results on the $16\times 8^3$ and $24\times
12^3$ lattices are striking (and was absent in the local determination). This finite-volume effect will be discussed in Sec.~\ref{sec:finitevolume}. We also
show for completeness the ratio $F_V/F_{PS}$ in Fig.~\ref{fvfpsplot}.

\section{Comments on finite-volume effects}
\label{sec:finitevolume}

\begin{figure}[!htp]
\centering
\includegraphics[scale=0.4]{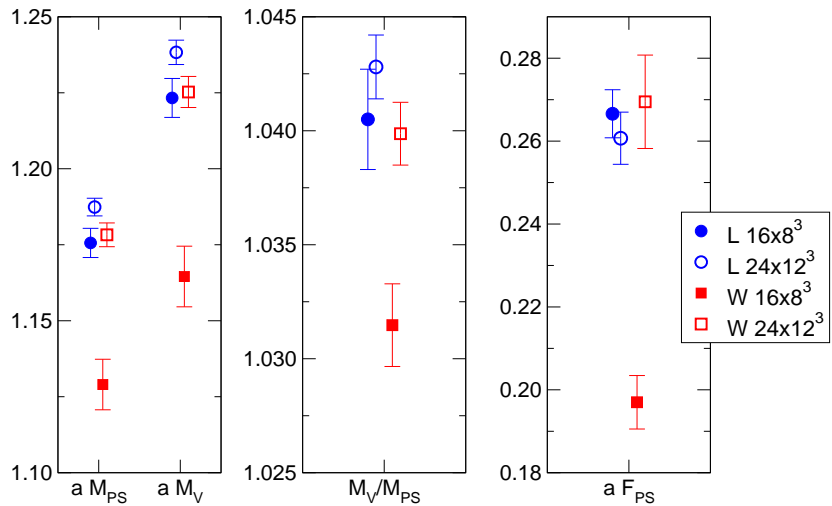}
\caption{PS and V masses, their ratio and PS decay constant on $16\times8^3$ and $24\times12^3$ lattices for $am_0 = -1.05$ and $\beta=2.25$ (L=local, W=wall).}
\label{fig:fvol}
\end{figure}

The wall-smeared results helped us to better understand how finite spatial volume affects the mesonic observables. In Fig.~\ref{fig:fvol} we plot the PS
and V masses, their ratio and PS decay constant on the $16\times8^3$ and $24\times12^3$ lattices for $am_0 = -1.05$ and $\beta=2.25$, both from local and
wall-smeared sources. For each observable, the gap between the two lattices becomes wider when wall-smeared sources are considered. Having only the data from
local sources, one can be tempted to underestimate the finite-volume errors. This would be a mistake: the mild dependence on the volume of the local data is
actually given by a cancellation of two larger effects: the finite volume and the bad determination of the plateaux in the effective masses.

\begin{figure}[!htp]
\centering
\includegraphics[scale=0.4]{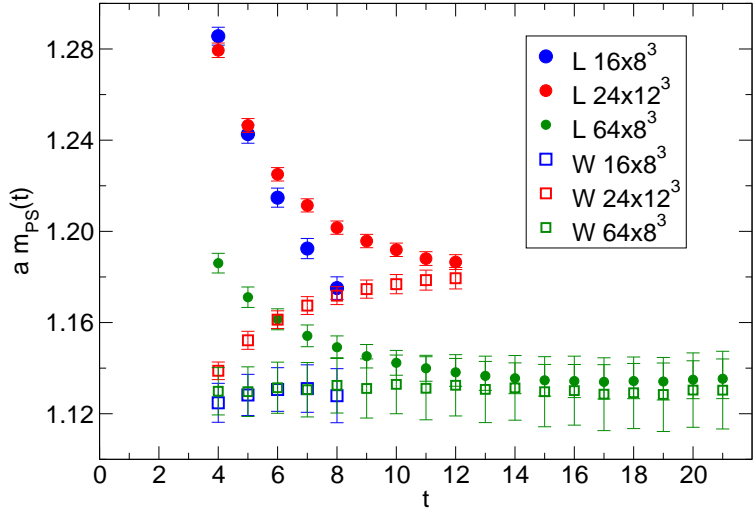}
\caption{Effective PS mass on different volumes for $am_0 = -1.05$ and $\beta=2.25$ (L=local, W=wall). At $t$ larger than $21$, this quantity (on the $64\times8^3$) becomes much noisier and we cut it for sake of clarity.}
\label{fig:fvol_mps}
\end{figure}

\begin{figure}[!htp]
\centering
\includegraphics[scale=0.4]{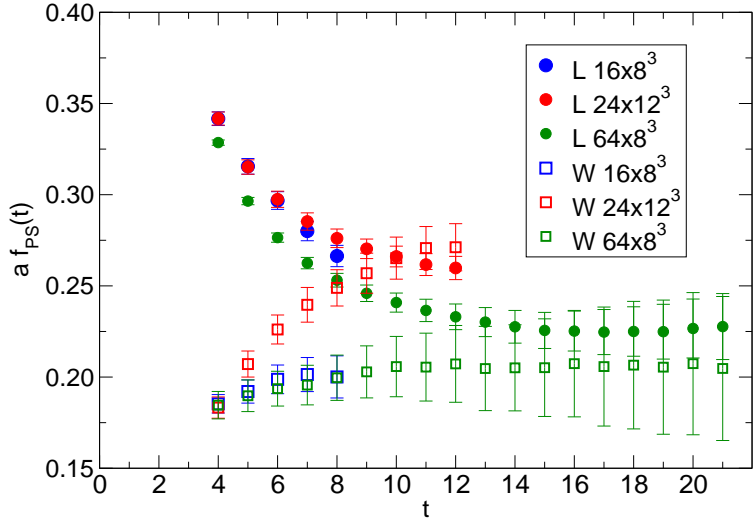}
\caption{Effective PS decay constant on different volumes for $am_0 = -1.05$ and $\beta=2.25$ (L=local, W=wall). At $t$ larger than $21$, this quantity (on the
$64\times8^3$) becomes much noisier and we cut it for the sake of clarity.}
\label{fig:fvol_fps}
\end{figure}

In order to clarify this point, it is useful to look directly at the effective PS mass (Fig.~\ref{fig:fvol_mps}) and the effective PS decay constant (Fig.~\ref{fig:fvol_fps}). We will comment on the effective PS mass, but all the observations will be equally valid for the effective PS decay constant.

The first observation is that the effective masses from local sources are always decreasing with the Euclidean time. Therefore, if the temporal size is not large enough to contain the plateau of the effective mass, the estimated mass will be larger than the real one. On the other hand the effective masses from wall-smeared sources for on this ensemble are increasing (although this is not true across all ensembles). Therefore, if the plateau is not reached, the estimated mass will be smaller than the real one.

Consider now the $24\times12^3$ effective masses in Fig.~\ref{fig:fvol_mps}. The local and wall-smeared sources give effective masses whose quality in terms of flatness is similar (compare with Fig.~\ref{eval12}), and the plateau is not clearly visible in any of the effective masses. However since the gap between the local and wall-smeared effective masses closes down in the midpoint $t=12$, one can argue that the plateau is effectively reached there.

The situation is completely different for the $16\times8^3$. The gap between the local and wall-smeared effective masses is always quite big. The wall-smeared source gives a much flatter effective mass than the local source (compare with Fig.~\ref{eval8}). In order to obtain a more precise estimate for the pseudoscalar mass on the spatial volume $8^3$, we simulated on a $64\times8^3$ lattice. In this case the temporal extent is large enough to obtain very good plateaux for both the local and wall-smeared effective masses.

By comparing the effective masses on the $24\times12^3$ and $64\times8^3$ lattices it is clear that the finite volume has the effect of making the pseudoscalar
meson lighter. What is happening then with the $16\times8^3$ lattice? The mass estimated with the local sources is affected by two relatively large effects: the
finite volume, which decreases the mass and the bad determination of the plateaux, which increases the mass. Having opposite sign and accidentally the same
magnitude, these two effects cancel each other. Therefore the finite volume effects are actually larger than what we estimated on the basis of the local
sources, and they are better estimated using the wall-smeared source at light enough masses.

The conclusions above are valid also for the vector meson mass and for the ratio $M_V/M_{PS}$. In particular from Fig.~\ref{mvmps_plot} it is clear that 
on increasing the spatial volume, the ratio $M_V/M_{PS}$ slightly increases, and this effect was completely hidden in the local-source determination.

\section{Conclusions}
\label{sec:conclusions}

In this article we have studied systematic effects on the PCAC mass, the mesonic masses and decay constants due to a short temporal size on the SU($2$) gauge theory with two Dirac fermions in the adjoint representation. In order to isolate the ground state in correlators one should take the source and sink infinitely distant. In practice one defines effective quantities (masses and decay constants) which depend on the time separation between source and sink, and which show a plateau at large distances. The value of the plateau gives an estimate for the corresponding mass or decay constant. At fixed temporal extent one can increase the relative amplitude of the ground state in correlators, using smeared sources and/or sinks. This translates into flatter and longer plateaux in the effective quantities.

We have extended the \emph{Chroma} suite of software in order to operate with fermions in the adjoint representation of the gauge group, and we have used the \emph{Chroma} built-in routines for measuring mesonic correlators with both Gaussian and wall-smeared sources. We observe that at our lightest masses the wall-smearing gives always the best overlap with the ground state. At heavy masses the mesonic wave functions are more localized and the local sources give a better overlap with the ground state. There is an intermediate regime of masses in which the local and wall-smearing sources yield plateaux of similar quality. In this case a Gaussian smearing with properly chosen width might be desirable. If one wants a procedure that enhance the overlap with the ground state at any mass, one should use a variational method with a large set of smeared sources. However, since the interesting physical region is close to the chiral limit, we chose simplicity against generality and we focused our detailed analysis on the wall-smearing only.

The enhancement of the plateaux with smeared sources does not come for free. Observables obtained with smeared sources have longer autocorrelation times. For a fixed set of configurations, a better control on the systematic error with respect to local sources is generally obtained at the cost of a larger statistical uncertainty.

Among the observables that we have considered, the PCAC mass is the least affected by the systematics, while the decay constants are the most affected. In the
region $a M_{PS} < 0.5$, the $16 \times 8^3$ lattice yields relative systematic errors for the PS mass larger than $10\%$. At least the $24 \times 12^3$ lattice
is needed in order to stay below $10\%$.

We also investigated how the finite temporal extent can conspire to partially mask effects due to finite spatial volume, and discovered that finite-volume effects were underestimated in our analysis with local sources. The relative difference between the determinations of the PS mass on the $16 \times 8^3$ and $24 \times 12^3$ lattices is of order $5\%$ at $aM_{PS} \simeq 1$ and it goes up to $14\%$ at about $aM_{PS} \simeq 0.3$. Again, in the interesting region of masses, the $16 \times 8^3$ lattice appears to be way too far from the infinite volume limit. A detailed study of finite-volume effects is extremely important in order to address issues like IR-conformality, and represents one of our major research lines.

Finally we notice that our conclusions regarding the near-conformal dynamics of this theory are robust, since the main qualitative features already presented in Refs.~\cite{DelDebbio:2009fd,DelDebbio:2010hu} are confirmed by the present analysis.

\section*{Acknowledgements}
The numerical calculations presented in this work have been performed on the Horseshoe6 cluster at the University of Southern Denmark (SDU) funded by the Danish
Centre for Scientific Computing for the project ``Origin of Mass'' 2009/2010. EK is supported by SUPA, the Scottish Universities Physics Alliance.  AR thanks
the Deutsche Forschungsgemeinschaft for financial support. BL is supported by the Royal Society. AP was supported by the EC (Research Infrastructure Action in
FP7, project \textit{HadronPhysics2}). The development of the code used in this work was partially supported by the EPSRC grant EP/F010303/1. We thank the DEISA
Consortium (www.deisa.eu), funded through the EU FP7 project RI-222919, for support within the DEISA Extreme Computing Initiative. LDD and BL would like to thank the GGI in Florence for hospitality during the final stage of this work.

\section*{\phantom{False}}
\begin{appendices}
\section{Correlators and smearings}
\label{app:smear}
\subsection{Local correlators}
In order to measure mesonic observables we measure zero-momentum correlators of the form
\begin{equation}
f_{\Gamma\Gamma^\prime}(t)=\sum_{\vec{x}}\langle
{\mathcal{O}^{SINK}_\Gamma}^\dagger(\vec{x},t)\mathcal{O}^{SRCE}_{\Gamma^\prime}(\vec{0},0)\rangle,\label{generalcorr}
\end{equation}
where $\mathcal{O}^{SRCE,SINK}_{\Gamma}$ are interpolating quark bilinear operators with the correct symmetries under spin and parity. We require the isospin
non-singlet correlators and so, for example, we could construct a \emph{local} correlator with the most immediate choice 
\begin{align}
 \mathcal{O}^{SRCE}_\Gamma(\vec{x},t)=\mathcal{O}^{SINK}_\Gamma(\vec{x},t)=\bar{\psi}_1(\vec{x},t)\Gamma\psi_2(\vec{x},t),\label{localops}
\end{align}
where the labels $i$ on the quark fields $\psi_i$ denotes the fermion flavour. Here $\Gamma$ is a matrix in the Dirac algebra, which
determines the symmetries of the operator.  This choice reproduces the correlators considered in \cite{DelDebbio:2010hu}:
\begin{equation}
f^{L}_{\Gamma\Gamma\prime}(t)=\sum_{\vec{x}}\langle\,\left(\,\bar{\psi}_1(\vec{x},t)\Gamma\psi_2(\vec{x},t)\,\right)^\dagger\,\bar{\psi}_1(\vec{0},
0)\Gamma^\prime\psi_2(\vec{0},0)\rangle, \label{localcorr}
\end{equation}
where here the superscript on $f_{\Gamma\Gamma^\prime}$ indicates the local choice. This correlator is measured by computing the quark propagator
$S(\vec{x},t;\vec{x}^\prime,t^\prime)$, in terms of which
\begin{equation}
 f^{L}_{\Gamma\Gamma^\prime}(t)=-\frac{a^3}{V_s}\sum_{\vec{x}}\mathrm{Tr}[\gamma_0\Gamma^\dagger\gamma_0S(\vec{x},t;\vec{0},0)\Gamma^\prime
\gamma_5S(\vec{x},t;\vec{0},0)^\dagger\gamma_5].
\end{equation}
The propagator is computed by solving the equation
\begin{equation}
a^4\sum_{\mathbf{y}}D(\mathbf{x};\mathbf{y})S(\mathbf{y},\mathbf{z})=I\delta_{\mathbf{x};\mathbf{z}},
\end{equation}
where the boldface variables denote the full space-time coordinate, $I$ denotes the identity matrix in spin and colour space, and $D(\mathbf{x};\mathbf{y})$
is the Dirac matrix.

\subsection{Extended quark fields}

\label{smearing}

In order to obtain an optimum signal for the masses we aim to extract from these correlators, we should construct interpolating operators with a maximised
overlap with the desired ground state. The local operators (\ref{localops}) are not expected to satisfy this requirement well, as the mesons typically have an
extension of many times the lattice spacing in a typical simulation. We can improve the situation by considering an operator which is extended spatially over
the lattice:
\begin{equation}
 \mathcal{O}_\Gamma(\vec{x},t)=\sum_{\vec{y}_1,\vec{y}_2}\Psi(\vec{x},\vec{y}_1,\vec{y}_2)\bar{\psi}_1(\vec{y}_1,t)\Gamma\psi_2(\vec{y}_2,t).
\end{equation}
Usually shell-model wave functions are used \cite{DeGrand:1990dz}, meaning the positions of the quark and antiquark are decoupled:
\begin{equation}
 \Psi(\vec{x},\vec{y}_1,\vec{y}_2)=\phi(\vec{x},\vec{y}_1)\phi(\vec{x},\vec{y}_2).\label{shellmodel}
\end{equation}
The choice $\phi(\vec{x},\vec{y})=\delta_{\vec{x},\vec{y}}$ reproduces the point-point case (\ref{localops}).

In general, such wave functions are not gauge invariant, and as such any expectation value over an ensemble of gauge configurations, in which they are used, must vanish, according to Elitzur's theorem \cite{Elitzur:1975im}. To avoid this we can fix the gauge on each configuration, being careful to check for errors introduced by the issue of Gribov copies.

Using $\phi^{SRCE/SINK}$ to define $\mathcal{O}^{SRCE/SINK}$ we see that our correlation function can be computed as
\begin{equation}
 f_{\Gamma\Gamma^\prime}(t)=-\sum_{\vec{x}}\mathrm{Tr}[\gamma_0\Gamma^\dagger\gamma_0\widehat{S}(\vec{x},t;\vec{0},0)\Gamma^\prime
\gamma_5\widehat{S}(\vec{x},t;\vec{0},0)^\dagger\gamma_5],
\end{equation}
where $\widehat{S}(\vec{x},t;\vec{x}^\prime,t^\prime)$ is defined as
\begin{equation}
\widehat{S}(\vec{x},t;\vec{x}^\prime,t^\prime)=\sum_{\vec{y},\vec{y}^\prime}S(\vec{y},t;\vec{y}^\prime,t^\prime)\phi^{SINK}(\vec{x},\vec{y})\phi^{SRCE}(\vec{x}
^\prime,\vec{y}^\prime).
\end{equation}
It can be easily seen that if we solve for $S^\prime$, the system
\begin{equation}
 a^4\sum_{\mathbf{y}}D(\mathbf{x},\mathbf{y})S^\prime(\mathbf{y},\mathbf{z})=\phi^{SRCE}(\vec{z},\vec{x})\delta_{x_0,z_0},
\end{equation}
we can compute $\widehat{S}$ as
\begin{equation}
 \widehat{S}(\vec{x},t;\vec{x}^\prime,t^\prime)=\sum_{\vec{y}}S^\prime(\vec{y},t;\vec{x}^\prime,t^\prime)\phi^{SINK}(\vec{x},\vec{y}).
\end{equation}
In fact it is the choice of a shell-model type wave-function (\ref{shellmodel}) that allows us to calculate the correlation function using only one inversion of the Dirac matrix (per colour and spin index).

\subsection{Smearing examples}

A simple guess for an effective form of $\phi(\vec{x})$ is in the form of a gaussian
\begin{equation}
\phi(\vec{x},\vec{y})=e^{-\left(\frac{|\vec{x}-\vec{y}|}{R}\right)^2},\label{gausssmear}
\end{equation}
where $R>0$ is some effective radius chosen to represent the wave function of the meson of interest. The choice $\frac{1}{R}\rightarrow0$ results in $\phi(\vec{x},\vec{y})$ having equal weight over the whole lattice, and is termed a \emph{wall smearing}.

On a lattice we can approximate the gaussian as the limit of the iterative form
\begin{equation}
 \phi(\vec{x},\vec{y})=\left(1-\frac{w^2}{4N}\Box\right)^N\delta_{\vec{x},\vec{y}},\label{lattgauss}
\end{equation}
where $\Box$ is the lattice version of the Laplacian
\begin{equation}
 \Box(\vec{x},\vec{y})=\sum_{i=1}^3\left(\delta_{\vec{x},\vec{y}-\hat{i}}+\delta_{\vec{x},\vec{y}+\hat{i}}\right).
\end{equation}
(\ref{lattgauss}) then approximates (\ref{gausssmear}) in the limit $N\rightarrow\infty$, with the radius $R$ being determined by $w$. Replacing $\Box$ with its covariant form
\begin{equation}
\Box(\vec{x},\vec{y};t)=\sum_{i=1}^3\left(U_i(\vec{x},t)\delta_{\vec{x},\vec{y}-\hat{i}}+U^\dagger_i(\vec{x}-\hat{i},t)\delta_{\vec{x},\vec{y}+\hat{i}}\right),
\end{equation}
results in a gauge invariant operator, negating the requirement for gauge fixing. This choice of $\phi$ is called \emph{gauge-invariant gaussian smearing}.

We have utilised both a wall-smearing (denoted $W$) and a gauge-invariant gaussian smearing (denoted $G$) in our study. 

\subsection{Gauge fixing}

When constructing a correlator involving the gauge-dependent wall-smeared quark bilinear, we must fix the gauge on each configuration with which we wish to
work. We fix to Coulomb gauge by generating a gauge-fixed gauge configuration from the original by maximising the
quantity $\displaystyle\sum_{\mathbf{x}}\sum_{i=1}^3\mathrm{Re}\left(\mathrm{Tr}[U_i(\mathbf{x})]\right)$.

\section{Meson correlator phenomenology}
\label{app:mesoncorr}

\subsection{Meson masses}
\label{mesonmass}
We extract the meson masses from our theory by analysing correlators of the form (\ref{generalcorr}) in the case where we consider source and sink operators
with equal symmetries, i.e. $\Gamma=\Gamma^\prime$, and so we shall write $f_{\Gamma\Gamma}=f_\Gamma$. We can write $f_\Gamma$ explicitly as an expectation
value on the vacuum state $|0\rangle$ :

\begin{equation}
 f_{\Gamma}(t)=\sum_{\vec{x}}\langle0| {\mathcal{O}^{SINK}_\Gamma}^\dagger(\vec{x},t)\mathcal{O}^{SRCE}_{\Gamma}(\vec{0},0)|0\rangle.
\end{equation}
Labelling the energy eigenstates of the theory as $|n,\vec{p}\rangle$, we can write a complete set of states as
\begin{equation}
\sum_n\int \frac{d^3p}{(2\pi)^32E_n(\vec{p})}\,|n,\vec{p}\rangle\langle n,\vec{p}|.
\end{equation}
We can insert this in $f_{\Gamma}$ producing
\begin{equation}
 f_{\Gamma}(t)=\sum_n\sum_{\vec{x}}\int \frac{d^3p}{(2\pi)^32E_n(\vec{p})}\langle0| {\mathcal{O}^{SINK}_\Gamma}^\dagger(\vec{x},t)|n,\vec{p}\rangle\langle
n,\vec{p}|\mathcal{O}^{SRCE}_{\Gamma}(\vec{0},0)|0\rangle.
\end{equation}
 Translating $\mathcal{O}^{SINK}(\mathbf{x})$ to the origin produces $e^{i\mathcal{P}\cdot\mathbf{x}}\mathcal{O}^{SINK}(\mathbf{0})e^{-i\mathcal{P}\cdot\mathbf{x}}$ where the four-momentum operator $\mathcal{P}=\{\mathcal{H},\vec{\mathcal{P}}\}$ giving
\begin{equation}
 \langle0| {\mathcal{O}^{SINK}_\Gamma}^\dagger(\vec{x},t)|n,\vec{p}\rangle=\langle0| {\mathcal{O}^{SINK}_\Gamma}^\dagger(\mathbf{0})|n,\vec{p}\rangle
e^{-i\mathbf{p}\cdot\mathbf{x}},
\end{equation}
where $\mathbf{p}=\{E_n(\vec{p}),\vec{p}\}$. As a result, the sum over the spatial position $\vec{x}$ collapses the sum onto zero-momentum
\begin{equation}
 f_{\Gamma}(t)=\sum_n\frac{1}{2E_n}\langle0| {\mathcal{O}^{SINK}_\Gamma}^\dagger(\vec{0},0)|n\rangle\langle n|\mathcal{O}^{SRCE}_{\Gamma}(\vec{0},0)|0\rangle
e^{-iE_nt},
\end{equation}
where we denote $|n,\vec{0}\rangle$ as $|n\rangle$ and $E_n(\vec{0})$ as $E_n$. The overlaps $\langle 0| \mathcal{O}(\mathbf{0})|n\rangle$ will vanish for all
states except those with the same symmetries as $\mathcal{O}_\Gamma$ and we can see that at large Euclidean time $\tau=it$ the correlator is dominated by the
lowest in energy of such states which we denote $|\Gamma\rangle$ with energy $E_\Gamma$ which as we are at zero momentum equals the mass of the state
$E_\Gamma=m_\Gamma$:
\begin{align}
 f_\Gamma(\tau)\stackrel{\tau\rightarrow\infty}{\longrightarrow}&\frac{1}{2m_\Gamma}\langle0| {\mathcal{O}^{SINK}_\Gamma}^\dagger(\vec{0},0)|\Gamma\rangle\langle \Gamma|\mathcal{O}^{SRCE}_{\Gamma}(\vec{0},0)|0\rangle e^{-m_\Gamma\tau}\notag\\
\equiv&A_\Gamma e^{-m_\Gamma\tau}.\label{corramps}
\end{align}
On a lattice with finite temporal extent $0<\tau<L_t$, this asymptotic behaviour is modified by the appearance of an extra term corresponding to a quark
propagating backward from source to sink through the anti-periodic boundary:
\begin{align}
  f_\Gamma(\tau)\rightarrow & A_\Gamma\left(e^{-m_\Gamma\tau}+e^{-m_\Gamma(L_t-\tau)}\right)\notag\\
\equiv& A_\Gamma hc(\tau,m_\Gamma,L_t).\label{finitet}
\end{align}
 
In this way we can extract the meson masses from the exponential behaviour of the $f_\Gamma$ at large Euclidean time. 



As in \cite{DelDebbio:2010hu}, we use the Prony method \cite{Fleming:2009wb} to solve this system, to produce an ``effective mass'' $m_\Gamma(\tau)$ which as a function of the lattice temporal
coordinate is expected to approach the desired mass in the limit of large times $m_\Gamma(t)\stackrel{\tau\rightarrow\infty}{\longrightarrow}m_\Gamma$. The
meson mass is extracted by choosing a region around the centre of the temporal axis and fitting the effective mass to a constant in this region. 

In our study we have considered the case $\Gamma=\gamma_5$, defining the \emph{pseudoscalar} channel, with mass $m_{\mathrm{PS}}$ and the degerate cases $\Gamma=\gamma_i$ $i\in\{1,2,3\}$, defining the \emph{vector} channel with mass $m_{\mathrm{V}}$. In practice the correlators $f_{\gamma_i}$ are averaged to produce a single correlator for the vector channel. We call the resulting vector correlator $f_{\mathrm{VV}}$ and the pseudoscalar correlator $f_{\mathrm{PP}}$. 

The masses can be extracted identically from these correlators regardless of the smearing used. In practice it is found that correlators with a smeared source
are preferred to local correlators for this purpose, in that they produce an improved signal to noise ratio for the masses. Correlators with smearing at
both the source and sink are found to be disfavoured because of enhanced fluctuations.

\subsection{Amplitudes}

If local quark fields are used, $\mathcal{O}^{SRCE/SINK}_\Gamma(\mathbf{x})=\bar{\psi}_1(\mathbf{x})\Gamma\psi_2(\mathbf{x})=\mathcal{O}^L_\Gamma(\mathbf{x})$.
In the case of both the pseudoscalar and vector channels, we are interested in the quantity $|\langle 0|\mathcal{O}^L_{\Gamma}(\mathbf{0})|\Gamma\rangle|$
although they have different meanings:
\begin{align}
 |\langle 0|\mathcal{O}^L_{\gamma_5}(\mathbf{0})|\gamma_5\rangle|\equiv&G_{\mathrm{PS}},\notag\\\label{ampeffops}\\
 |\langle 0|\mathcal{O}^L_{\gamma_i}(\mathbf{0})|\gamma_i\rangle|\equiv&\epsilon_iF_{\mathrm{V}}m_{\mathrm{V}},\notag
\end{align}
where $\epsilon_i$ is a polarisation tensor. We call $G_{\mathrm{PS}}$ the \emph{psuedoscalar vacuum to meson amplitude} (or, more commonly, simply the
psuedoscalar amplitude), and $F_{\mathrm{V}}$ is the \emph{vector decay constant}. We can easily construct effective observables for these quantities from the
local correlators $f^{L}_{PP}$ and $f^{L}_{VV}$:
\begin{align}
 G^L_{\mathrm{PS}}(\tau)=&\sqrt{\frac{2m_{\mathrm{PS}}(\tau)f^{L}_{\mathrm{PP}}(\tau)}{hc(\tau,m_{\mathrm{PS}}(\tau),L_t)}},\notag\\\label{localamps}\\
F^L_{\mathrm{V}}(\tau)=&\sqrt{\frac{2f^{L}_{\mathrm{VV}}(\tau)}{m_{\mathrm{V}}hc(\tau,m_{\mathrm{V}}(\tau),L_t)}}.\notag
\end{align}
If we wish to use smeared operators to extract these quantities, the amplitudes in (\ref{corramps}) are, in general,  no longer related to the quantities
of interest (\ref{ampeffops}). However, if our correlator involves only a smearing at the source, with a local sink, we see that the sink amplitude in
(\ref{corramps}) is still of the correct form (\ref{ampeffops}). We need 
 cancel the other undesired amplitude, introduced by the smearing. We can do this by combining our local-smeared correlator ($f^{LS}_\Gamma$) with a
smeared-smeared correlator ($f^{SS}_\Gamma$). Effective observables equivalent to (\ref{ampeffops}) can be defined from smeared correlators as 
\begin{align}
G^S_{\mathrm{PS}}(\tau)=&\sqrt{\frac{2m_{\mathrm{PS}}(\tau)}{hc(\tau,m_{\mathrm{PS}}(\tau),L_t)}\frac{{f^{LS}_{\mathrm{PP}}}^2(\tau)}{f^{SS}_{\mathrm{PP}}(\tau)
}},\notag\\\label
{smearedamps}\\
F^S_{\mathrm{V}}(\tau)=&\sqrt{\frac{2}{m_{\mathrm{V}}hc(\tau,m_{\mathrm{V}}(\tau),L_t)}\frac{{f^{LS}_{\mathrm{VV}}}^2(\tau)}{f^{SS}_{\mathrm{VV}}(\tau)}}.\notag
\end{align}

 \subsection{Quark Mass}
As our simulation is based on the Wilson quark formulation, the physical quark mass in our simulation $m$ is related to the bare quark mass which is an input to
the simulation $m_0$ by an additive renormalisation, which being a non-perturbative quantity can not be calculated a priori. As such we must have a method of
determining the physical quark mass in the simulation in order to determine our proximity to the chiral point $m=0$ and to observe the scaling of mesonic
observables with $m$. 

The most straight-forward such method is via the \emph{partially conserved axial current mass} or PCAC mass. We define the continuum non-singlet axial and pseudoscalar currents as 
\begin{align}
 A_\mu(\mathbf{x})=\bar{\psi}_1(\mathbf{x})\gamma_\mu\gamma_5\psi_2(\mathbf{x}),&&P(\mathbf{x})=\bar{\psi}_1(\mathbf{x})\gamma_5\psi_2(\mathbf{x}).
\end{align}
We see that these are continuum versions of our $\mathcal{O}^L_{\gamma_\mu\gamma_5}$ and $\mathcal{O}^L_{\gamma_5}$. From the Ward identity for the axial
transformation $\psi\rightarrow e^{i\alpha\gamma_5}\psi$ we obtain for the divergence of the axial current
\begin{equation}
 \partial_\mu A_\mu(\mathbf{x})=-2mP(\mathbf{x}),
\end{equation}
where $m$ is the physical quark mass, as above. From this we obtain
\begin{equation}
 \frac{\partial}{\partial t }\int d^3x\,\langle A_0(\vec{x},t)\mathcal{O}_{\gamma_5}\rangle=-2m\langle
P(\vec{x},t)\mathcal{O}_{\gamma_5}\rangle,\label{timeward}
\end{equation}
where $\mathcal{O}_{\gamma_5}$ is any bilinear quark operator with the symmetries of a pseudoscalar current. Taking a lattice version of this, and choosing for $\mathcal{O}_{\gamma_5}$ any of the local or smeared lattice pseudoscalar currents we have previously constructed, we see we can define an effective PCAC quark mass via
\begin{equation}
 m(\tau)=\frac{m_{\mathrm{PS}}}{\sinh(am_{\mathrm{PS}})}\frac{f^{LS}_{\mathrm{AP}}(\tau-a)-f^{LS}_{\mathrm{AP}}(\tau+a)}{4f^{LS}_{\mathrm{PP}}(\tau)},\label{
effpcac}
\end{equation}
where we define $f_{\mathrm{AP}}$ to be $f_{\gamma_0\gamma_5,\gamma_5}$. The prefactor of $\frac{m_{\mathrm{PS}}}{\sinh(am_{\mathrm{PS}})}$ arises by a choice of the lattice finite difference operator which more accurately represents the continuum derivative on $f_{\mathrm{AP}}$. The correlators $f^{LS}$ are constructed with a local sink, and a source which can be local, or involve any smearing. 

\subsection{Pseudoscalar decay constant}

Similarly to (\ref{corramps}) the correlator $f_{\mathrm{AP}}$ has an asymptotic behaviour:
\begin{align}
 f_{\mathrm{AP}} (\tau)\stackrel{\tau\rightarrow\infty}{\longrightarrow}&\frac{1}{2m_{\mathrm{PS}}}\langle0| {\mathcal{O}^{SINK}_{\gamma_0\gamma_5}}^\dagger(\vec{0},0)|\gamma_5\rangle\langle \gamma_5|\mathcal{O}^{SRCE}_{\gamma_5}(\vec{0},0)|0\rangle e^{-m_{\mathrm{PS}}\tau}\notag\\
\equiv&A_{\mathrm{AP}}e^{-m_{\mathrm{PS}}\tau}.
\end{align}
In contrast to (\ref{finitet}) however, the contribution to $f_{\mathrm{AP}}$ from propagation around the lattice comes with the opposite sign, so on a lattice
with finite temporal extent,
\begin{align}
 f_{\mathrm{AP}}(\tau)\rightarrow& A_{\mathrm{AP}}\left(e^{-m_{\mathrm{PS}}\tau}-e^{-m_{\mathrm{PS}}(L_t-\tau)}\right)\notag\\
\equiv &A_{\mathrm{AP}}hs(\tau,m_{\mathrm{PS}},L_t).
\end{align}
Now we define the \emph{pseudoscalar decay constant}  $F_\mathrm{PS}$ as
\begin{equation}
 m_\mathrm{PS}F_\mathrm{PS}=\langle0| \mathcal{O}^L_{\gamma_0\gamma_5}(\vec{0},0)|\gamma_5\rangle.
\end{equation}
Combining this with the Ward identity for $f_{\mathrm{AP}}$ we can define an effective observable for $F_{\mathrm{PS}}$ as
\begin{equation}
 F^S_{\mathrm{PS}}(\tau)=\frac{2m(\tau)G^S_{\mathrm{PS}}(\tau)}{m_{\mathrm{PS}}^2(\tau)}.\label{efffps}
\end{equation}
The superscript $S$ here indicates that this is valid for observables obtained from any smeared correlator, provided the corresponding definition of $G_{\mathrm{PS}}$ is used, from (\ref{localamps}) or (\ref{smearedamps}).

\section{Results tables}
\label{app:results}

\begin{table}[!htp]
 \centering
\input{final_res_table_16x8x8x8b2.25.tex}
\caption{Results for mesonic observables from wall-smeared correlators on a $16\times8^3$ lattice at $\beta=2.25$.}
\label{restable16}
\end{table}
\begin{table}[!htp]
 \centering
 \input{final_res_table_24x12x12x12b2.25.tex}
\caption{Results for mesonic observables from wall-smeared correlators on a $24\times12^3$ lattice at $\beta=2.25$.}
\label{restable24}
\end{table}
\begin{table}[!htp]
 \centering
 \input{final_res_table_32x16x16x16b2.25.tex}
\caption{Results for mesonic observables from wall-smeared correlators on a $32\times16^3$ lattice at $\beta=2.25$.}
\label{restable32}
\end{table}
\begin{table}[!htp]
 \centering
 \input{final_res_table_64x24x24x24b2.25.tex}
\caption{Results for mesonic observables from wall-smeared correlators on a $64\times24^3$ lattice at $\beta=2.25$.}
\label{restable64_1}
\end{table}
\begin{table}[!htp]
 \centering
 \input{final_res_table_64x24x24x24b2.1.tex}
\caption{Results for mesonic observables from wall-smeared correlators on a $64\times24^3$ lattice at $\beta=2.1$.}
\label{restable64_2}
\end{table}

\begin{table}[!htp]
 \centering
\input{final_rat_table_16x8x8x8b2.25.tex}
\caption{Ratios of mesonic observables from wall-smeared correlators on a $16\times8^3$ lattice at $\beta=2.25$.}
\label{rattable16}
\end{table}
\begin{table}[!htp]
 \centering
 \input{final_rat_table_24x12x12x12b2.25.tex}
\caption{Ratios of mesonic observables from wall-smeared correlators on a $24\times12^3$ lattice at $\beta=2.25$.}
\label{rattable24}
\end{table}
\begin{table}[!htp]
 \centering
 \input{final_rat_table_32x16x16x16b2.25.tex}
\caption{Ratios of mesonic observables from wall-smeared correlators on a $32\times16^3$ lattice at $\beta=2.25$.}
\label{rattable32}
\end{table}
\begin{table}[!htp]
 \centering
 \input{final_rat_table_64x24x24x24b2.25.tex}
\caption{Ratios of mesonic observables from wall-smeared correlators on a $64\times24^3$ lattice at $\beta=2.25$.}
\label{rattable64_1}
\end{table}
\begin{table}[!htp]
 \centering
 \input{final_rat_table_64x24x24x24b2.1.tex}
\caption{Ratios of mesonic observables from wall-smeared correlators on a $64\times24^3$ lattice at $\beta=2.1$.}
\label{rattable64_2}
\end{table}

\section{Pull tables}
\label{app:pull}

\input{pull_res_table_16x8x8x8b2.25.tex}
 \input{pull_res_table_24x12x12x12b2.25.tex}
 \input{pull_res_table_32x16x16x16b2.25.tex}
 \input{pull_res_table_64x24x24x24b2.25.tex}

\input{pull_rat_table_16x8x8x8b2.25.tex}
 \input{pull_rat_table_24x12x12x12b2.25.tex}
 \input{pull_rat_table_32x16x16x16b2.25.tex}
 \input{pull_rat_table_64x24x24x24b2.25.tex}

\end{appendices}

\bibliographystyle{JHEP}
\bibliography{refs}

\end{document}

%% file: final_res_table_16x8x8x8b2.25.tex
\begin{tabular}{cccccccc}
\hline
lattice & $-am_0$ & $N_{conf}$ & $am$ & $am_{PS}$ & $am_V$ & $aF_{PS}$ & $aF_V$ \\
\hline
S0 & -0.5 & 901
 & 1.16353(73) & 2.7983(15) & 2.8042(16) & 0.2950(73) & 0.3338(92)  \\
S1 & -0.25 & 901
 & 1.07205(97) & 2.6535(21) & 2.6613(22) & 0.3150(63) & 0.3629(80)  \\
S2 & 0 & 901
 & 0.9706(11) & 2.4938(25) & 2.5045(27) & 0.3335(63) & 0.3935(84)  \\
S3 & 0.25 & 901
 & 0.8552(11) & 2.3092(28) & 2.3241(31) & 0.3579(74) & 0.435(10)  \\
S4 & 0.5 & 901
 & 0.7224(13) & 2.0934(32) & 2.1155(37) & 0.3729(87) & 0.475(13)  \\
S5 & 0.75 & 901
 & 0.5607(18) & 1.8136(47) & 1.8473(55) & 0.375(12) & 0.511(21)  \\
S6 & 0.9 & 901
 & 0.4330(18) & 1.5582(68) & 1.5987(81) & 0.315(13) & 0.441(23)  \\
A0 & 0.95 & 1501
 & 0.3849(16) & 1.4488(68) & 1.4902(84) & 0.291(12) & 0.411(22)  \\
A1 & 0.975 & 1499
 & 0.3582(17) & 1.3830(74) & 1.4251(91) & 0.274(11) & 0.388(21)  \\
A2 & 1 & 7300
 & 0.3314(19) & 1.3137(78) & 1.3553(97) & 0.258(13) & 0.368(23)  \\
A3 & 1.025 & 1481
 & 0.3001(19) & 1.2222(90) & 1.260(11) & 0.230(10) & 0.324(18)  \\
A4 & 1.05 & 1481
 & 0.2688(15) & 1.1290(83) & 1.1645(99) & 0.1970(64) & 0.2692(89)  \\
A5 & 1.075 & 1277
 & 0.2352(18) & 1.011(13) & 1.042(16) & 0.185(11) & 0.258(20)  \\
A6 & 1.1 & 1279
 & 0.1992(32) & 0.886(14) & 0.914(19) & 0.1642(99) & 0.227(17)  \\
A7 & 1.125 & 1344
 & 0.1595(25) & 0.725(14) & 0.747(18) & 0.1478(76) & 0.200(15)  \\
A8 & 1.15 & 1278
 & 0.1150(31) & 0.519(18) & 0.534(23) & 0.1439(70) & 0.194(10)  \\
A9 & 1.175 & 1280
 & 0.0628(30) & 0.285(23) & 0.295(30) & 0.1569(43) & 0.2120(81)  \\
\end{tabular}

%% file: final_res_table_24x12x12x12b2.25.tex
\begin{tabular}{cccccccc}
\hline
lattice & $-am_0$ & $N_{conf}$ & $am$ & $am_{PS}$ & $am_V$ & $aF_{PS}$ & $aF_V$ \\
\hline
B0 & 0.95 & 1973
 & 0.39017(68) & 1.4720(23) & 1.5186(28) & 0.3220(61) & 0.468(11)  \\
B1 & 1 & 1689
 & 0.33623(82) & 1.3441(28) & 1.3932(37) & 0.2942(73) & 0.434(13)  \\
B2 & 1.05 & 1564
 & 0.27470(91) & 1.1782(39) & 1.2252(51) & 0.269(11) & 0.395(19)  \\
B3 & 1.075 & 1438
 & 0.2393(10) & 1.0660(55) & 1.1058(68) & 0.231(13) & 0.333(24)  \\
B4 & 1.1 & 5112
 & 0.2014(10) & 0.9310(65) & 0.9638(79) & 0.1523(85) & 0.208(14)  \\
B5 & 1.125 & 1240
 & 0.16013(92) & 0.7697(60) & 0.7963(68) & 0.1485(63) & 0.2062(97)  \\
B6 & 1.15 & 640
 & 0.1149(15) & 0.572(12) & 0.588(15) & 0.0955(53) & 0.1296(89)  \\
B7 & 1.175 & 5137
 & 0.0653(14) & 0.3277(95) & 0.336(11) & 0.0985(33) & 0.1298(54)  \\
B8 & 1.18 & 818
 & 0.0547(17) & 0.282(11) & 0.294(13) & 0.0984(53) & 0.1311(78)  \\
B9 & 1.185 & 840
 & 0.0418(16) & 0.206(11) & 0.213(11) & 0.1056(43) & 0.1366(63)  \\
B10 & 1.19 & 700
 & 0.0300(11) & 0.1476(82) & 0.1539(96) & 0.1129(32) & 0.1496(36)  \\
\end{tabular}

%% file: final_res_table_32x16x16x16b2.25.tex
\begin{tabular}{cccccccc}
\hline
lattice & $-am_0$ & $N_{conf}$ & $am$ & $am_{PS}$ & $am_V$ & $aF_{PS}$ & $aF_V$ \\
\hline
C0 & 1.15 & 1090
 & 0.11731(77) & 0.6121(64) & 0.6305(83) & 0.0983(44) & 0.1314(87)  \\
C1 & 1.175 & 523
 & 0.06579(77) & 0.3652(87) & 0.381(10) & 0.0746(35) & 0.1037(64)  \\
C2 & 1.18 & 917
 & 0.05437(79) & 0.3042(69) & 0.3174(80) & 0.0736(37) & 0.0992(59)  \\
C3 & 1.185 & 864
 & 0.04217(84) & 0.2241(62) & 0.2297(72) & 0.0763(32) & 0.0992(44)  \\
C4 & 1.19 & 1083
 & 0.03065(72) & 0.1682(62) & 0.1764(64) & 0.0776(27) & 0.1038(36)  \\
\end{tabular}

%% file: final_res_table_64x24x24x24b2.25.tex
\begin{tabular}{cccccccc}
\hline
lattice & $-am_0$ & $N_{conf}$ & $am$ & $am_{PS}$ & $am_V$ & $aF_{PS}$ & $aF_V$ \\
\hline
D0 & 1.18 & 185
 & 0.05528(25) & 0.3239(49) & 0.3295(62) & 0.0398(27) & 0.0455(41)  \\
D1 & 1.185 & 164
 & 0.04287(29) & 0.2462(58) & 0.2566(75) & 0.0696(65) & 0.091(11)  \\
D2 & 1.19 & 160
 & 0.02967(50) & 0.1741(52) & 0.1759(59) & 0.0501(30) & 0.0601(52)  \\
\end{tabular}

%% file: final_res_table_64x24x24x24b2.1.tex
\begin{tabular}{cccccccc}
\hline
lattice & $-am_0$ & $N_{conf}$ & $am$ & $am_{PS}$ & $am_V$ & $aF_{PS}$ & $aF_V$ \\
\hline
E0 & 1.25 & 131
 & 0.11751(28) & 0.7173(11) & 0.7735(39) & 0.1592(26) & 0.263(12)  \\
E1 & 1.26 & 130
 & 0.08527(34) & 0.5612(15) & 0.5881(45) & 0.1122(26) & 0.169(11)  \\
\end{tabular}

%% file: final_rat_table_16x8x8x8b2.25.tex
\begin{tabular}{cccccccc}
\hline
lattice & $-am_0$ & $am_{PS}^2/m$ & $m_V/F_{PS}$ & $m_V/m_{PS}$ & $a^3 (m_{PS}F_{PS})^2/m$ & $F_V/F_{PS}$\\
\hline
S0 & -0.5 & 6.7300(39) & 9.50(23) & 1.002100(47) & 0.586(29) & 1.1311(35)  \\
S1 & -0.25 & 6.5681(57) & 8.45(16) & 1.00294(10) & 0.652(26) & 1.1520(29)  \\
S2 & 0 & 6.4070(67) & 7.51(13) & 1.00428(14) & 0.713(27) & 1.1797(37)  \\
S3 & 0.25 & 6.2347(86) & 6.49(13) & 1.00647(22) & 0.798(33) & 1.2172(50)  \\
S4 & 0.5 & 6.065(10) & 5.67(12) & 1.01058(36) & 0.844(40) & 1.2754(72)  \\
S5 & 0.75 & 5.866(17) & 4.91(16) & 1.01858(75) & 0.830(57) & 1.359(14)  \\
S6 & 0.9 & 5.606(31) & 5.07(19) & 1.0259(11) & 0.558(49) & 1.398(18)  \\
A0 & 0.95 & 5.453(33) & 5.11(19) & 1.0286(13) & 0.465(41) & 1.408(19)  \\
A1 & 0.975 & 5.338(37) & 5.20(20) & 1.0303(15) & 0.402(37) & 1.414(18)  \\
A2 & 1 & 5.206(40) & 5.24(24) & 1.0316(17) & 0.350(36) & 1.421(24)  \\
A3 & 1.025 & 4.977(47) & 5.47(22) & 1.0308(19) & 0.265(26) & 1.405(21)  \\
A4 & 1.05 & 4.741(48) & 5.91(16) & 1.0314(18) & 0.184(13) & 1.366(18)  \\
A5 & 1.075 & 4.352(94) & 5.63(29) & 1.0300(50) & 0.150(21) & 1.390(32)  \\
A6 & 1.1 & 3.94(10) & 5.58(27) & 1.0310(70) & 0.107(14) & 1.385(37)  \\
A7 & 1.125 & 3.299(93) & 5.06(20) & 1.0307(70) & 0.0724(87) & 1.358(54)  \\
A8 & 1.15 & 2.35(13) & 3.71(17) & 1.028(16) & 0.0489(60) & 1.350(31)  \\
A9 & 1.175 & 1.30(19) & 1.88(20) & 1.034(44) & 0.0320(48) & 1.351(39)  \\
\end{tabular}

%% file: final_rat_table_24x12x12x12b2.25.tex
\begin{tabular}{cccccccc}
\hline
lattice & $-am_0$ & $am_{PS}^2/m$ & $m_V/F_{PS}$ & $m_V/m_{PS}$ & $a^3 (m_{PS}F_{PS})^2/m$ & $F_V/F_{PS}$\\
\hline
B0 & 0.95 & 5.553(10) & 4.717(84) & 1.03166(59) & 0.576(22) & 1.454(12)  \\
B1 & 1 & 5.373(13) & 4.73(11) & 1.03652(96) & 0.465(23) & 1.476(20)  \\
B2 & 1.05 & 5.053(21) & 4.55(17) & 1.0398(13) & 0.367(31) & 1.465(19)  \\
B3 & 1.075 & 4.748(35) & 4.79(26) & 1.0373(21) & 0.255(30) & 1.438(31)  \\
B4 & 1.1 & 4.303(42) & 6.34(32) & 1.0352(18) & 0.100(11) & 1.365(27)  \\
B5 & 1.125 & 3.700(42) & 5.36(20) & 1.0345(18) & 0.0818(77) & 1.388(14)  \\
B6 & 1.15 & 2.84(10) & 6.17(26) & 1.0287(85) & 0.0261(35) & 1.355(38)  \\
B7 & 1.175 & 1.644(71) & 3.41(12) & 1.026(12) & 0.0160(14) & 1.317(25)  \\
B8 & 1.18 & 1.457(93) & 2.99(19) & 1.042(21) & 0.0141(18) & 1.332(30)  \\
B9 & 1.185 & 1.022(90) & 2.02(12) & 1.034(23) & 0.0114(12) & 1.293(44)  \\
B10 & 1.19 & 0.727(68) & 1.36(10) & 1.043(32) & 0.00926(75) & 1.325(37)  \\
\end{tabular}

%% file: final_rat_table_32x16x16x16b2.25.tex
\begin{tabular}{cccccccc}
\hline
lattice & $-am_0$ & $am_{PS}^2/m$ & $m_V/F_{PS}$ & $m_V/m_{PS}$ & $a^3 (m_{PS}F_{PS})^2/m$ & $F_V/F_{PS}$\\
\hline
C0 & 1.15 & 3.194(53) & 6.41(25) & 1.0299(47) & 0.0310(30) & 1.334(40)  \\
C1 & 1.175 & 2.028(87) & 5.12(21) & 1.045(10) & 0.0113(14) & 1.389(37)  \\
C2 & 1.18 & 1.702(62) & 4.32(20) & 1.043(10) & 0.0092(10) & 1.348(28)  \\
C3 & 1.185 & 1.191(55) & 3.01(13) & 1.024(16) & 0.00696(68) & 1.298(23)  \\
C4 & 1.19 & 0.924(57) & 2.27(11) & 1.048(20) & 0.00557(46) & 1.336(26)  \\
\end{tabular}

%% file: final_rat_table_64x24x24x24b2.25.tex
\begin{tabular}{cccccccc}
\hline
lattice & $-am_0$ & $am_{PS}^2/m$ & $m_V/F_{PS}$ & $m_V/m_{PS}$ & $a^3 (m_{PS}F_{PS})^2/m$ & $F_V/F_{PS}$\\
\hline
D0 & 1.18 & 1.899(55) & 8.29(45) & 1.0169(88) & 0.00304(51) & 1.140(47)  \\
D1 & 1.185 & 1.414(64) & 3.70(28) & 1.042(14) & 0.0069(15) & 1.308(59)  \\
D2 & 1.19 & 1.023(53) & 3.51(17) & 1.009(19) & 0.00259(42) & 1.199(54)  \\
\end{tabular}

%% file: final_rat_table_64x24x24x24b2.1.tex
\begin{tabular}{cccccccc}
\hline
lattice & $-am_0$ & $am_{PS}^2/m$ & $m_V/F_{PS}$ & $m_V/m_{PS}$ & $a^3 (m_{PS}F_{PS})^2/m$ & $F_V/F_{PS}$\\
\hline
E0 & 1.25 & 4.378(10) & 4.857(81) & 1.0783(49) & 0.1110(37) & 1.653(76)  \\
E1 & 1.26 & 3.693(21) & 5.24(12) & 1.0479(79) & 0.0465(22) & 1.50(10)  \\
\end{tabular}

%% file: pull_res_table_16x8x8x8b2.25.tex
\begin{table}
\centering
\begin{tabular}{ccccccccc}
\hline
lattice & V &  $-am_0$ & $am$ & $am_{PS}$ & $a^2 G_{PS}$ & $aF_{PS}$ & $am_{V}$ &  $ a F_{V}$ \\
\hline
S0  & $16\times 8^3$ & -0.5 & 0.357743 & 0.630827 & 0.235108 & 0.256243 & 0.620554 & 0.183502  \\
S1  & $16\times 8^3$ & -0.25 & 0.0890535 & 0.37089 & 0.25073 & 0.185906 & 0.364952 & 0.221968  \\
S2  & $16\times 8^3$ & 0 & 0.176656 & 0.103129 & 0.633081 & 0.570268 & 0.117478 & 0.603322  \\
S3  & $16\times 8^3$ & 0.25 & 0.0544221 & 0.89148 & 0.546006 & 0.402862 & 0.863709 & 0.524572  \\
S4  & $16\times 8^3$ & 0.5 & 0.41261 & 0.213066 & 0.74581 & 0.769548 & 0.198067 & 0.685742  \\
S5  & $16\times 8^3$ & 0.75 & 0.798726 & 0.721695 & 0.723448 & 0.773543 & 0.76452 & 0.658796  \\
S6  & $16\times 8^3$ & 0.9 & 0.914745 & 2.12396 & 2.85272 & 2.62616 & 2.19736 & 2.96989  \\
A0  & $16\times 8^3$ & 0.95 & 1.17047 & 2.87797 & 3.53652 & 3.17306 & 3.0289 & 3.70567  \\
A1  & $16\times 8^3$ & 0.975 & 1.48267 & 3.03748 & 3.0976 & 3.69649 & 3.1175 & 4.28469  \\
A2  & $16\times 8^3$ & 1 & 1.08558 & 3.30753 & 3.8366 & 3.39684 & 3.38991 & 3.99593  \\
A3  & $16\times 8^3$ & 1.025 & 1.50436 & 4.02795 & 5.56733 & 4.93876 & 4.08713 & 5.90623  \\
A4  & $16\times 8^3$ & 1.05 & 1.44732 & 4.8367 & 9.42537 & 8.00532 & 4.94294 & 12.0426  \\
A5  & $16\times 8^3$ & 1.075 & 0.854806 & 3.47175 & 6.88026 & 4.07907 & 3.40687 & 5.06541  \\
A6  & $16\times 8^3$ & 1.1 & 0.761814 & 3.25019 & 5.45758 & 4.42701 & 3.03732 & 5.34694  \\
A7  & $16\times 8^3$ & 1.125 & 1.07367 & 3.4146 & 5.77455 & 4.51791 & 3.05006 & 4.97578  \\
A8  & $16\times 8^3$ & 1.15 & 0.769772 & 2.64252 & 3.77974 & 2.61207 & 2.05669 & 3.82454  \\
A9  & $16\times 8^3$ & 1.175 & 0.567182 & 1.70204 & 1.91251 & 0.632631 & 1.61141 & 2.25052  \\
\end{tabular}
\caption{Pull of wall-smeared results from local results on $16\times 8^3$ lattice.\label{respulls16x8x8x8b2.25}}
\end{table}

%% file: pull_res_table_24x12x12x12b2.25.tex
\begin{table}
\centering
\begin{tabular}{ccccccccc}
\hline
lattice & V &  $-am_0$ & $am$ & $am_{PS}$ & $a^2 G_{PS}$ & $aF_{PS}$ & $am_{V}$ &  $ a F_{V}$ \\
\hline
B0  & $24\times 12^3$ & 0.95 & 0.774467 & 0.783603 & 1.42361 & 1.40895 & 0.871324 & 1.53757  \\
B1  & $24\times 12^3$ & 1 & 0.159875 & 1.40115 & 1.18323 & 0.807577 & 1.36064 & 1.46937  \\
B2  & $24\times 12^3$ & 1.05 & 0.457775 & 1.85951 & 0.417244 & 0.680213 & 2.01302 & 0.207503  \\
B3  & $24\times 12^3$ & 1.075 & 0.431929 & 2.35926 & 0.376053 & 0.0367313 & 2.57414 & 0.556833  \\
B4  & $24\times 12^3$ & 1.1 & 0.27378 & 4.08446 & 5.1113 & 4.65501 & 4.10009 & 5.44129  \\
B5  & $24\times 12^3$ & 1.125 & 0.0929123 & 4.42596 & 3.32675 & 1.76942 & 3.94236 & 2.57502  \\
B6  & $24\times 12^3$ & 1.15 & 0.886415 & 2.49618 & 4.85751 & 3.77226 & 2.04429 & 3.86822  \\
B7  & $24\times 12^3$ & 1.175 & 0.252409 & 2.90899 & 3.77684 & 1.8568 & 2.29301 & 2.32682  \\
B8  & $24\times 12^3$ & 1.18 & 0.623981 & 1.97381 & 2.98903 & 1.74813 & 1.6552 & 2.72508  \\
B9  & $24\times 12^3$ & 1.185 & 0.45537 & 1.69178 & 2.27698 & 0.544327 & 0.994413 & 2.10543  \\
B10  & $24\times 12^3$ & 1.19 & 0.107576 & 1.61624 & 1.86186 & 0.8264 & 0.831428 & 1.55631  \\
\end{tabular}
\caption{Pull of wall-smeared results from local results on $24\times 12^3$ lattice.\label{respulls24x12x12x12b2.25}}
\end{table}

%% file: pull_res_table_32x16x16x16b2.25.tex
\begin{table}
\centering
\begin{tabular}{ccccccccc}
\hline
lattice & V &  $-am_0$ & $am$ & $am_{PS}$ & $a^2 G_{PS}$ & $aF_{PS}$ & $am_{V}$ &  $ a F_{V}$ \\
\hline
C0  & $32\times 16^3$ & 1.15 & 0.0826998 & 2.76528 & 4.43289 & 3.49837 & 2.51094 & 3.69838  \\
C1  & $32\times 16^3$ & 1.175 & 0.656922 & 1.80444 & 3.88521 & 2.7899 & 1.63612 & 3.34912  \\
C2  & $32\times 16^3$ & 1.18 & 0.297786 & 2.34619 & 3.68642 & 2.42467 & 2.03804 & 3.61078  \\
C3  & $32\times 16^3$ & 1.185 & 0.0677226 & 2.36342 & 2.47807 & 0.443416 & 1.66751 & 1.68151  \\
C4  & $32\times 16^3$ & 1.19 & 0.13612 & 2.19063 & 3.10032 & 0.7232 & 1.75468 & 2.79363  \\
\end{tabular}
\caption{Pull of wall-smeared results from local results on $32\times 16^3$ lattice.\label{respulls32x16x16x16b2.25}}
\end{table}

%% file: pull_res_table_64x24x24x24b2.25.tex
\begin{table}
\centering
\begin{tabular}{ccccccccc}
\hline
lattice & V &  $-am_0$ & $am$ & $am_{PS}$ & $a^2 G_{PS}$ & $aF_{PS}$ & $am_{V}$ &  $ a F_{V}$ \\
\hline
D0  & $64\times 24^3$ & 1.18 & 0.231276 & 1.68437 & 10.7124 & 3.65489 & 1.98021 & 2.99979  \\
D1  & $64\times 24^3$ & 1.185 & 0.753454 & 0.491626 & 8.31795 & 2.83008 & 0.281183 & 1.97354  \\
D2  & $64\times 24^3$ & 1.19 & 0.302909 & 0.0376987 & 10.0776 & 0.153963 & 0.124777 & 0.183138  \\
\end{tabular}
\caption{Pull of wall-smeared results from local results on $64\times 24^3$ lattice.\label{respulls64x24x24x24b2.25}}
\end{table}

%% file: pull_rat_table_16x8x8x8b2.25.tex
\begin{table}
\centering
\begin{tabular}{cccccccc}
\hline
lattice & V & $-am_0$ & $am_{PS}^2/m$ & $m_V/F_{PS}$ & $m_V/m_{PS}$ & $a^3 (m_{PS}F_{PS})^2/m$ & $F_V/F_{PS}$ \\
\hline
S0 & $16\times 8^3$ & -0.5 & 579.839 & 103.336 & 42.4374 & 14.8134 & 29.6658  \\
S1 & $16\times 8^3$ & -0.25 & 355.964 & 115.643 & 28.3108 & 17.5745 & 40.2534  \\
S2 & $16\times 8^3$ & 0 & 256.472 & 102.845 & 27.2575 & 17.0714 & 36.0592  \\
S3 & $16\times 8^3$ & 0.25 & 161.5 & 76.8603 & 27.1434 & 13.7659 & 30.298  \\
S4 & $16\times 8^3$ & 0.5 & 105.789 & 55.6038 & 25.8027 & 10.1067 & 25.9244  \\
S5 & $16\times 8^3$ & 0.75 & 67.0073 & 41.9002 & 22.8086 & 6.02327 & 18.2194  \\
S6 & $16\times 8^3$ & 0.9 & 162.403 & 326 & 21.9192 & 10.0395 & 20.4498  \\
A0 & $16\times 8^3$ & 0.95 & 42.936 & 27.5523 & 19.6178 & 1.61548 & 14.8302  \\
A1 & $16\times 8^3$ & 0.975 & 34.6138 & 20.3632 & 18.5986 & 5.08223 & 15.0627  \\
A2 & $16\times 8^3$ & 1 & 29.2933 & 12.2385 & 15.9358 & 7.94612 & 10.4417  \\
A3 & $16\times 8^3$ & 1.025 & 27.2297 & 8.32147 & 13.3442 & 16.2042 & 10.4521  \\
A4 & $16\times 8^3$ & 1.05 & 27.2413 & 2.95847 & 13.634 & 32.67 & 7.68135  \\
A5 & $16\times 8^3$ & 1.075 & 16.7288 & 0.223305 & 3.82008 & 26.723 & 3.43554  \\
A6 & $16\times 8^3$ & 1.1 & 16.2186 & 2.76273 & 1.6808 & 24.4004 & 0.595224  \\
A7 & $16\times 8^3$ & 1.125 & 19.8065 & 2.29941 & 0.273288 & 21.8378 & 1.64239  \\
A8 & $16\times 8^3$ & 1.15 & 21.8515 & 4.29543 & 0.285916 & 27.8008 & 3.78825  \\
A9 & $16\times 8^3$ & 1.175 & 20.4856 & 11.9002 & 0.00540329 & 27.7572 & 3.35479  \\
\end{tabular}
\caption{Pull of wall-smeared results from local results on $16\times 8^3$ lattice.\label{ratpulls16x8x8x8b2.25}}
\end{table}

%% file: pull_rat_table_24x12x12x12b2.25.tex
\begin{table}
\centering
\begin{tabular}{cccccccc}
\hline
lattice & V & $-am_0$ & $am_{PS}^2/m$ & $m_V/F_{PS}$ & $m_V/m_{PS}$ & $a^3 (m_{PS}F_{PS})^2/m$ & $F_V/F_{PS}$ \\
\hline
B0 & $24\times 12^3$ & 0.95 & 0.396346 & 1.38681 & 0.71794 & 1.40066 & 0.62282  \\
B1 & $24\times 12^3$ & 1 & 0.495822 & 0.681721 & 0.7434 & 0.948186 & 1.25164  \\
B2 & $24\times 12^3$ & 1.05 & 0.562277 & 0.944498 & 1.53568 & 0.612448 & 2.12925  \\
B3 & $24\times 12^3$ & 1.075 & 1.16716 & 0.223536 & 1.73031 & 0.172471 & 1.35859  \\
B4 & $24\times 12^3$ & 1.1 & 2.54333 & 3.71056 & 2.36472 & 5.76623 & 2.85406  \\
B5 & $24\times 12^3$ & 1.125 & 3.11413 & 1.03624 & 0.533927 & 2.64676 & 1.60002  \\
B6 & $24\times 12^3$ & 1.15 & 1.51009 & 3.61068 & 0.122117 & 4.28713 & 0.420703  \\
B7 & $24\times 12^3$ & 1.175 & 3.05379 & 0.169209 & 0.492669 & 3.59039 & 0.854043  \\
B8 & $24\times 12^3$ & 1.18 & 1.84539 & 0.456036 & 0.226258 & 2.9395 & 1.20918  \\
B9 & $24\times 12^3$ & 1.185 & 1.71852 & 0.384545 & 1.35089 & 1.91835 & 1.48933  \\
B10 & $24\times 12^3$ & 1.19 & 1.74255 & 1.0647 & 0.897999 & 1.21423 & 1.90816  \\
\end{tabular}
\caption{Pull of wall-smeared results from local results on $24\times 12^3$ lattice.\label{ratpulls24x12x12x12b2.25}}
\end{table}

%% file: pull_rat_table_32x16x16x16b2.25.tex
\begin{table}
\centering
\begin{tabular}{cccccccc}
\hline
lattice & V & $-am_0$ & $am_{PS}^2/m$ & $m_V/F_{PS}$ & $m_V/m_{PS}$ & $a^3 (m_{PS}F_{PS})^2/m$ & $F_V/F_{PS}$ \\
\hline
C0 & $32\times 16^3$ & 1.15 & 1.93171 & 3.06861 & 0.882135 & 4.34396 & 1.47396  \\
C1 & $32\times 16^3$ & 1.175 & 1.06468 & 2.32532 & 0.137237 & 3.31804 & 0.411924  \\
C2 & $32\times 16^3$ & 1.18 & 2.03676 & 1.54644 & 0.0619247 & 3.39686 & 1.57362  \\
C3 & $32\times 16^3$ & 1.185 & 2.39644 & 0.563872 & 0.653246 & 1.82322 & 1.42152  \\
C4 & $32\times 16^3$ & 1.19 & 2.26197 & 0.559476 & 0.623971 & 2.59077 & 1.87611  \\
\end{tabular}
\caption{Pull of wall-smeared results from local results on $32\times 16^3$ lattice.\label{ratpulls32x16x16x16b2.25}}
\end{table}

%% file: pull_rat_table_64x24x24x24b2.25.tex
\begin{table}
\centering
\begin{tabular}{cccccccc}
\hline
lattice & V & $-am_0$ & $am_{PS}^2/m$ & $m_V/F_{PS}$ & $m_V/m_{PS}$ & $a^3 (m_{PS}F_{PS})^2/m$ & $F_V/F_{PS}$ \\
\hline
D0 & $64\times 24^3$ & 1.18 & 2.04515 & 3.80398 & 1.32766 & 3.07595 & 1.80012  \\
D1 & $64\times 24^3$ & 1.185 & 0.739296 & 3.8496 & 0.192169 & 2.02015 & 0.0468305  \\
D2 & $64\times 24^3$ & 1.19 & 0.139092 & 0.267849 & 0.156939 & 0.100296 & 0.348643  \\
\end{tabular}
\caption{Pull of wall-smeared results from local results on $64\times 24^3$ lattice.\label{ratpulls64x24x24x24b2.25}}
\end{table}

%% file: imp_spec_mwt.bbl
\providecommand{\href}[2]{#2}\begingroup\raggedright\begin{thebibliography}{10}

\bibitem{Weinberg:1975gm}
S.~Weinberg, {\it {Implications of Dynamical Symmetry Breaking}},  {\em Phys.
  Rev.} {\bf D13} (1976) 974--996.

\bibitem{Susskind:1978ms}
L.~Susskind, {\it {Dynamics of Spontaneous Symmetry Breaking in the Weinberg-
  Salam Theory}},  {\em Phys. Rev.} {\bf D20} (1979) 2619--2625.

\bibitem{Holdom:1984sk}
B.~Holdom, {\it {Techniodor}},  {\em Phys. Lett.} {\bf B150} (1985) 301.

\bibitem{Yamawaki:1985zg}
K.~Yamawaki, M.~Bando, and K.-i. Matumoto, {\it {Scale Invariant Technicolor
  Model and a Technidilaton}},  {\em Phys. Rev. Lett.} {\bf 56} (1986) 1335.

\bibitem{Appelquist:1986an}
T.~W. Appelquist, D.~Karabali, and L.~C.~R. Wijewardhana, {\it {Chiral
  Hierarchies and the Flavor Changing Neutral Current Problem in Technicolor}},
   {\em Phys. Rev. Lett.} {\bf 57} (1986) 957.

\bibitem{Luty:2004ye}
M.~A. Luty and T.~Okui, {\it {Conformal technicolor}},  {\em JHEP} {\bf 09}
  (2006) 070, [\href{http://xxx.lanl.gov/abs/hep-ph/0409274}{{\tt
  hep-ph/0409274}}].

\bibitem{Hill:2002ap}
C.~T. Hill and E.~H. Simmons, {\it {Strong dynamics and electroweak symmetry
  breaking}},  {\em Phys. Rept.} {\bf 381} (2003) 235--402,
  [\href{http://xxx.lanl.gov/abs/hep-ph/0203079}{{\tt hep-ph/0203079}}].

\bibitem{Sannino:2009za}
F.~Sannino, {\it {Conformal Dynamics for TeV Physics and Cosmology}},  {\em
  Acta Phys. Polon.} {\bf B40} (2009) 3533--3743,
  [\href{http://xxx.lanl.gov/abs/0911.0931}{{\tt arXiv:0911.0931}}].

\bibitem{Piai:2010ma}
M.~Piai, {\it {Lectures on walking technicolor, holography and gauge/gravity
  dualities}},  {\em Adv. High Energy Phys.} {\bf 2010} (4302)
  [\href{http://xxx.lanl.gov/abs/1004.0176}{{\tt arXiv:1004.0176}}].

\bibitem{Nunez:2008wi}
C.~Nunez, I.~Papadimitriou, and M.~Piai, {\it {Walking Dynamics from String
  Duals}},  {\em Int. J. Mod. Phys.} {\bf A25} (2010) 2837--2865,
  [\href{http://xxx.lanl.gov/abs/0812.3655}{{\tt arXiv:0812.3655}}].

\bibitem{Elander:2009pk}
D.~Elander, C.~Nunez, and M.~Piai, {\it {A light scalar from walking solutions
  in gauge-string duality}},  {\em Phys. Lett.} {\bf B686} (2010) 64--67,
  [\href{http://xxx.lanl.gov/abs/0908.2808}{{\tt arXiv:0908.2808}}].

\bibitem{DeGrand:2010ba}
T.~DeGrand, {\it {Lattice studies of QCD-like theories with many fermionic
  degrees of freedom}},  \href{http://xxx.lanl.gov/abs/1010.4741}{{\tt
  arXiv:1010.4741}}.

\bibitem{DelDebbio:2010zz}
L.~Del~Debbio, {\it {The conformal window on the lattice}},  {\em PoS} {\bf
  LATTICE2010} (2010) 004.

\bibitem{Dietrich:2006cm}
D.~D. Dietrich and F.~Sannino, {\it {Walking in the SU(N)}},  {\em Phys. Rev.}
  {\bf D75} (2007) 085018, [\href{http://xxx.lanl.gov/abs/hep-ph/0611341}{{\tt
  hep-ph/0611341}}].

\bibitem{Hietanen:2009az}
A.~J. Hietanen, K.~Rummukainen, and K.~Tuominen, {\it {Evolution of the
  coupling constant in SU(2) lattice gauge theory with two adjoint fermions}},
  {\em Phys. Rev.} {\bf D80} (2009) 094504,
  [\href{http://xxx.lanl.gov/abs/0904.0864}{{\tt arXiv:0904.0864}}].

\bibitem{Bursa:2009we}
F.~Bursa, L.~Del~Debbio, L.~Keegan, C.~Pica, and T.~Pickup, {\it {Mass
  anomalous dimension in SU(2) with two adjoint fermions}},  {\em Phys. Rev.}
  {\bf D81} (2010) 014505, [\href{http://xxx.lanl.gov/abs/0910.4535}{{\tt
  arXiv:0910.4535}}].

\bibitem{DeGrand:2011qd}
T.~DeGrand, Y.~Shamir, and B.~Svetitsky, {\it {Infrared fixed point in SU(2)
  gauge theory with adjoint fermions}},
  \href{http://xxx.lanl.gov/abs/1102.2843}{{\tt arXiv:1102.2843}}.

\bibitem{Catterall:2007yx}
S.~Catterall and F.~Sannino, {\it {Minimal walking on the lattice}},  {\em
  Phys. Rev.} {\bf D76} (2007) 034504,
  [\href{http://xxx.lanl.gov/abs/0705.1664}{{\tt arXiv:0705.1664}}].

\bibitem{DelDebbio:2008zf}
L.~Del~Debbio, A.~Patella, and C.~Pica, {\it {Higher representations on the
  lattice: numerical simulations. SU(2) with adjoint fermions}},  {\em Phys.
  Rev.} {\bf D81} (2010) 094503, [\href{http://xxx.lanl.gov/abs/0805.2058}{{\tt
  arXiv:0805.2058}}].

\bibitem{Catterall:2008qk}
S.~Catterall, J.~Giedt, F.~Sannino, and J.~Schneible, {\it {Phase diagram of
  SU(2) with 2 flavors of dynamical adjoint quarks}},  {\em JHEP} {\bf 11}
  (2008) 009, [\href{http://xxx.lanl.gov/abs/0807.0792}{{\tt
  arXiv:0807.0792}}].

\bibitem{Hietanen:2009zz}
A.~Hietanen, J.~Rantaharju, K.~Rummukainen, and K.~Tuominen, {\it {Minimal
  technicolor on the lattice}},  {\em Nucl. Phys.} {\bf A820} (2009)
  191c--194c.

\bibitem{DelDebbio:2009fd}
L.~Del~Debbio, B.~Lucini, A.~Patella, C.~Pica, and A.~Rago, {\it {Conformal
  versus confining scenario in SU(2) with adjoint fermions}},  {\em Phys.Rev.}
  {\bf D80} (2009) 074507, [\href{http://xxx.lanl.gov/abs/0907.3896}{{\tt
  arXiv:0907.3896}}].

\bibitem{Catterall:2009sb}
S.~Catterall, J.~Giedt, F.~Sannino, and J.~Schneible, {\it {Probes of nearly
  conformal behavior in lattice simulations of minimal walking technicolor}},
  \href{http://xxx.lanl.gov/abs/0910.4387}{{\tt arXiv:0910.4387}}.

\bibitem{DelDebbio:2010hx}
L.~Del~Debbio, B.~Lucini, A.~Patella, C.~Pica, and A.~Rago, {\it {The infrared
  dynamics of Minimal Walking Technicolor}},  {\em Phys. Rev.} {\bf D82} (2010)
  014510, [\href{http://xxx.lanl.gov/abs/1004.3206}{{\tt arXiv:1004.3206}}].

\bibitem{DelDebbio:2010hu}
L.~Del~Debbio, B.~Lucini, A.~Patella, C.~Pica, and A.~Rago, {\it {Mesonic
  spectroscopy of Minimal Walking Technicolor}},  {\em Phys.Rev.} {\bf D82}
  (2010) 014509, [\href{http://xxx.lanl.gov/abs/1004.3197}{{\tt
  arXiv:1004.3197}}].

\bibitem{DeGrand:2009mt}
T.~DeGrand and A.~Hasenfratz, {\it {Remarks on lattice gauge theories with
  infrared-attractive fixed points}},  {\em Phys. Rev.} {\bf D80} (2009)
  034506, [\href{http://xxx.lanl.gov/abs/0906.1976}{{\tt arXiv:0906.1976}}].

\bibitem{Lucini:2009an}
B.~Lucini, {\it {Strongly Interacting Dynamics beyond the Standard Model on a
  Spacetime Lattice}},  \href{http://xxx.lanl.gov/abs/0911.0020}{{\tt
  arXiv:0911.0020}}.

\bibitem{DelDebbio:2010jy}
L.~Del~Debbio and R.~Zwicky, {\it {Scaling relations for the entire spectrum in
  mass-deformed conformal gauge theories}},
  \href{http://xxx.lanl.gov/abs/1009.2894}{{\tt arXiv:1009.2894}}.

\bibitem{DelDebbio:2010ze}
L.~Del~Debbio and R.~Zwicky, {\it {Hyperscaling relations in mass-deformed
  conformal gauge theories}},  {\em Phys. Rev.} {\bf D82} (2010) 014502,
  [\href{http://xxx.lanl.gov/abs/1005.2371}{{\tt arXiv:1005.2371}}].

\bibitem{Kerrane:2010xq}
E.~Kerrane {\em et.~al.}, {\it {Improved Spectroscopy of Minimal Walking
  Technicolor}},  \href{http://xxx.lanl.gov/abs/1011.0607}{{\tt
  arXiv:1011.0607}}.

\bibitem{Edwards:2004sx}
{\bf SciDAC Collaboration, LHPC Collaboration, UKQCD Collaboration}
  Collaboration, R.~G. Edwards and B.~Joo, {\it {The Chroma software system for
  lattice QCD}},  {\em Nucl.Phys.Proc.Suppl.} {\bf 140} (2005) 832,
  [\href{http://xxx.lanl.gov/abs/hep-lat/0409003}{{\tt hep-lat/0409003}}].

\bibitem{Madras:1988ei}
N.~Madras and A.~D. Sokal, {\it {The Pivot algorithm: a highly efficient Monte
  Carlo method for selfavoiding walk}},  {\em J.Statist.Phys.} {\bf 50} (1988)
  109--186.

\bibitem{DeGrand:1990dz}
T.~A. DeGrand and R.~D. Loft, {\it {Wave function tests for lattice QCD
  spectroscopy}},  {\em Comput.Phys.Commun.} {\bf 65} (1991) 84--91.

\bibitem{Elitzur:1975im}
S.~Elitzur, {\it {Impossibility of Spontaneously Breaking Local Symmetries}},
  {\em Phys.Rev.} {\bf D12} (1975) 3978--3982.

\bibitem{Fleming:2009wb}
G.~T. Fleming, S.~D. Cohen, H.-W. Lin, and V.~Pereyra, {\it {Excited-State
  Effective Masses in Lattice QCD}},  {\em Phys. Rev.} {\bf D80} (2009) 074506,
  [\href{http://xxx.lanl.gov/abs/0903.2314}{{\tt arXiv:0903.2314}}].

\end{thebibliography}\endgroup
